\documentclass[twocolumn]{revtex4}

\usepackage{amsfonts}
\usepackage{amsmath}
\usepackage{amssymb}

\usepackage{ragged2e}
\usepackage{graphicx}
\usepackage[usenames,dvipsnames]{color}
\definecolor{darkblue}{RGB}{0,0,196}

\begin{document}

\title{Observation of different scenarios in different temperatures in small and large collision systems \vspace{0.5cm}}

\author{Muhammad Waqas$^{1}$\footnote{waqas\_phy313@yahoo.com, waqas\_phy313@ucas.ac.cn},
Lu-Meng Liu$^{2}$\footnote{liulumeng18@mails.ucas.ac.cn},
Guang-Xiong Peng$^{1,3,4}$\footnote{Correspondence: gxpeng@ucas.edu.ac.cn},
Muhammad Ajaz$^{5}$\footnote{Correspondence: ajaz@awkum.edu.pk, muhammad.ajaz@cern.ch} \vspace{0.25cm}}

\affiliation{$^1$ School of Nuclear Science and Technology, University of Chinese Academy of Sciences,
Beijing 100049, People's Republic of China
\\
$^2$ School of Physical Sciences, University of Chinese Academy of Sciences, Beijing 100049, China \\
$^3$ Theoretical Physics Center for Science Facilities, Institute of High Energy Physics, Beijing 100049, China
\\
$^4$ Synergetic Innovation Center for Quantum Effects \& Applications, Hunan Normal University, Changsha 410081, China
\\
$^5$ Department of Physics, Abdul Wali Khan University Mardan, 23200 Mardan, Pakistan}

\begin{abstract}

\vspace{0.5cm}

\noindent {\bf Abstract:} We used the modified Hagedron function and analyzed the experimental data measured by
the BRAHMS, STAR, PHENIX and ALICE Collaborations in Copper-Copper, Gold-Gold, deuteron-Gold, Lead-Lead, proton-Lead
and proton-proton collisions, and extracted the related parameters (kinetic freeze-out temperature, transverse flow
velocity, kinetic freeze-out volume, mean transverse momentum and initial temperature) from the transverse momentum
spectra of the particles (non-strange and strange particles). We observed that all the above parameters
decrease from central to peripheral collisions, except transverse flow velocity which remains unchanged from central to
peripheral collisions. The kinetic freeze-out temperature depends on the cross-section interaction of the particle such that
larger cross-section of the particle corresponds to smaller $T_0$, and reveals the two kinetic freeze-out scenario, while
the initial temperature depends on the mass of the particle and it increase with the particle mass. The transverse flow velocity
and mean transverse momentum depends on the mass of the particle and the former decrease while the later increase with the particle mass. In addition, the kinetic kinetic freeze-out
volume also decrease with particle mass which reveals the volume differential freeze-out scenario and indicates different
 freeze-out surfaces for different particles. We also extracted the entropy index-parameter
$n$ and the parameter $N_0$, and the former remains almost unchanged while the later decrease from central to peripheral collisions.
Furthermore, the kinetic freeze-out temperature, transverse flow velocity, kinetic freeze-out volume, initial temperature, mean
transverse momentum and the parameter $N_0$ at LHC are larger than that of RHIC, and they show their dependence on the collision cross-section as well as on collision energy at RHIC and LHC.
\\

\end{abstract}

\maketitle

\section{Introduction}

A suitable tool for the production of the hot and dense matter namely Quark-Gluon Plasma (QGP) in the
laboratory \cite{1,2} is the high energy collisions. This state of matter is formed in the initial stages
of collision, and it survives for a short period of time (7-10 fm/c) and then rapidly transforms into
a hadron gas system. Because of the multi-partonic interactions throughout the evolution of the
collision, the information about the initial condition of the system gets lost. In order to obtain the
final state behavior of such a colliding system, we need the measurement of number and identity of the
produced particles along with their energy and momentum spectra. The final state information are very useful to understand the production mechanism of particles and the nature of produced matter in
these high energy collisions.

The chemical and kinetic freeze-out are the two symbolic freeze-out conditions where the space-time
evolution of the fireball produced in the collision cease. The colliding medium first reaches
chemical equilibrium, where the inelastic scattering stops due to expansion of the system which
results in the stabilization of the particle chemistry in the fireball. This stage is called
chemical freeze-out stage and the temperature at this stage is known as chemical freeze-out temperature
$(T_{ch})$ \cite{4}. The relative fractions of the particles are fixed but they still interact with each other,
and then stops until the final state interactions between them are no longer effective. This stage is
called thermal/kinetic freeze-out stage and the temperature at this stage is known as kinetic freeze-out
temperature \cite{5,6,7,8}. In Standard model of heavy ion collisions, the kinetic freeze-out occurs after the chemical
freeze-out due to large mean free path of the later which is also claimed in ref. \cite{9}. Generally, the
transverse momentum ($p_T$) spectra as well as the yields of the produced hadrons constitute some basic
measurements for the extraction of parameters of chemical and kinetic freeze-out.

A multiple chemical freeze-out scenario with the early fixing of chemical composition of strange hadron
compared to non-strange light hadrons is advocated in ref. \cite{10}. The early chemical freeze-out of strange
hadrons is due to their small inelastic cross-section. A natural question arise in mind whether a similar
hierarchy structure also occurs at kinetic decoupling, or the mass dependent hierarchy
occurs in case of kinetic freeze-out because the variation in medium induced momentum for heavy hadrons
would be smaller than the lighter hadrons and therefore, with the decrease of temperature of the fireball,
the earlier kinetic decoupling of heavy hadrons is expected.

In the present article, we will analyze the bulk properties in terms of kinetic freeze-out temperature ($T_0$),
transverse flow velocity ($\beta_T$) and kinetic freeze-out volume ($V$). All these mentioned parameters are
discussed in detail in various literatures \cite{11,12,13,14}. In the present work, we choose different collision systems
such as small and large systems in order to check the dependence of the above parameters on the size of
interacting system and different particles are chosen in order to check the differences in different particle
emissions.

Before going to the next section, we would like to point out that we also analyse the initial temperature which
occurs in the initial stage of collisions and which is also important to study because
the phenomenon in the initial stages of collisions and the freeze-out stages is different.

The remainder of the paper consists of formalism and method, results discussions and conclusions.

\section{Formalism and method}

The structure of transverse momentum spectra ($p_T$) of the charged particles is very complex,
and therefore is distributed into various regions. Especially, when the $p_T$ range approaches
to 100 GeV/c at LHC collisions \cite{1a}. There are different $p_T$regions according to the model
analysis \cite{2a}. The first $p_T$ region include $p_T$$<$4-6 GeV/c and second region include 4-6 GeV/c$<$ $p_T$$<$ 17-20 GeV/c, while the third region includes $p_T$$>$17-20 GeV/c. It is believed that different
$p_T$ regions (soft, hard and very hard $p_T$ region) signify different interacting mechanisms.
There are different explanations due to different models and methods even for the same $p_T$ region. In the
present work, the maximum $p_T$ range is 7-8 GeV/c, therefore we will be limited to the soft and hard
process and will skip the very hard process. There is a very soft region also which will be also skipped and it corresponds to $p_T$ $<$0.5 GeV/c which involves the resonance production that is not described by the model, but one can read ref. \cite{3a,4a} for more details of very soft and very hard processes. Generally, the soft excitation process is
distributed in a narrow $p_T$ range of less than 2-3 GeV/c or a little more, and mostly light flavor particles
are produced in this region.
The soft process have many choices of formalisms such as Blast wave model with Boltzmann Gibb's
statistics \cite{5a,6a,7a,8a}, Blast wave model with Tsallis statistics \cite{9a,10a,11a,12a},Tsallis pareto-type function \cite{12aa}, Erlang distribution \cite{13a,14a,15a},
Scwinger mechanism \cite{16a,17a,18a,19a}, Hagedorn distribution function \cite{20a}, Standard distribution \cite{21a} etc. The $p_T$ region above 3 GeV/c is
contributed for the hard process and is described by Quantum chromodynamics (QCD) calculus \cite{22a,23a,24a}
or inverse power law which is also known as Hagedron function

\begin{align}
f_0(p_T)=&\frac{1}{N}\frac{\mathrm{d}N}{\mathrm{d}p_\mathrm{T}}= Ap_T \bigg( 1+\frac{p_T}{p_0} \bigg)^{-n},
\end{align}
where $A$, is the normalization constant, and $p_0$ and $n$ are free parameters. This function is revised
in different forms in \cite{25a,26a,27a,28a,29a,30a} and each has different significance.

In the present work, we use the Hagedorn function with embeded transverse flow velocity \cite{31a}, which can be
expressed as
\begin{align}
f_0(p_T)=&\frac{1}{N}\frac{\mathrm{d}N}{\mathrm{d}p_\mathrm{T}}= 2\pi C p_T \bigg[1+<\gamma_t> \nonumber\\
\frac{(m_T-p_T<\beta_T>)}{n T_0}\bigg]^{-n}
\end{align}
where $C$ stands for the normalization constant that leads the integral in Eq. (2) to be normalized to 1,
$m_T$ is the transverse mass and is equal to $\sqrt{p_T^2+m_0^2}$, $m_0$ is the rest mass of the particle,
In Eq. 2, $C$=$gV/(2\pi)^2$. So Eq. 2 becomes as
\begin{align}
f_0(p_T)=&\frac{1}{N}\frac{\mathrm{d}N}{\mathrm{d}p_\mathrm{T}}= \frac{gV}{2\pi}  p_T \bigg[1+<\gamma_t> \nonumber\\
\frac{(m_T-p_T<\beta_T>)}{nT_0}\bigg]^{-n}
\end{align}

where $g$ is the degeneracy factor which differs for every particle, depends on
their spin based on $g_n$=2$S_n$+1.

\begin{figure*}[htbp]
\begin{center}
\includegraphics[width=11.0cm]{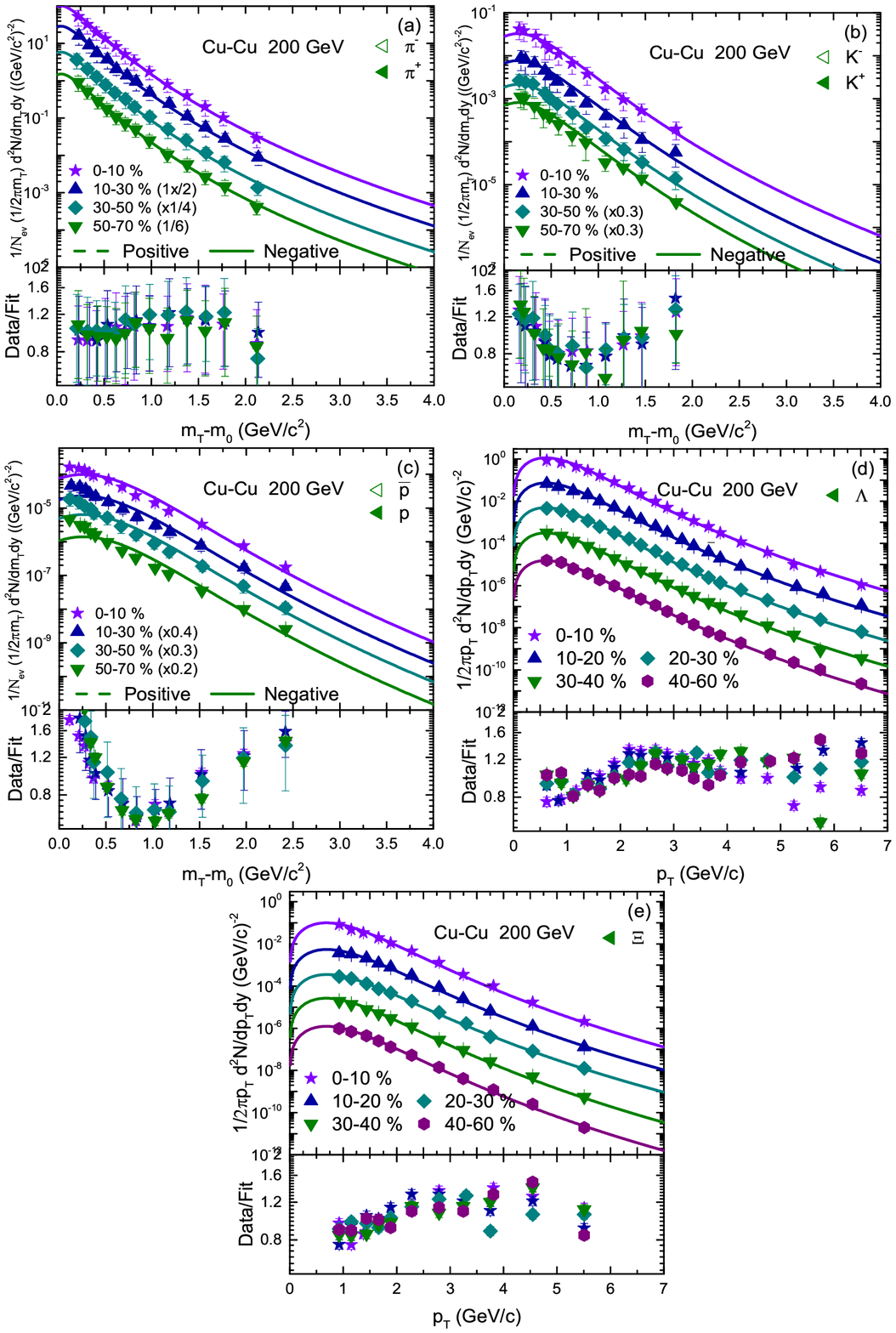}
\end{center}
\justifying\noindent {Fig. 1. The transverse momentum spectra of $\pi^+$, $\pi^-$, $K^+$, $K^-$, $p$,
$\bar p$, $\Lambda$ and $\Xi$ in Copper-Copper (Cu-Cu) collisions at 200 GeV in various centrality classes. The experimental data of BRAHMS Collaboration at RHIC for $\pi^+$, $\pi^-$, $K^+$, $K^-$, $p$ and $\bar p$ is taken from ref. \cite{1c} at $|y|<$0, while for $\Lambda$ and $\Xi$ is from ref. \cite{2c} at $|y|<$0.5.}
\end{figure*}

\begin{figure*}[htbp]
\begin{center}
\includegraphics[width=11.0cm]{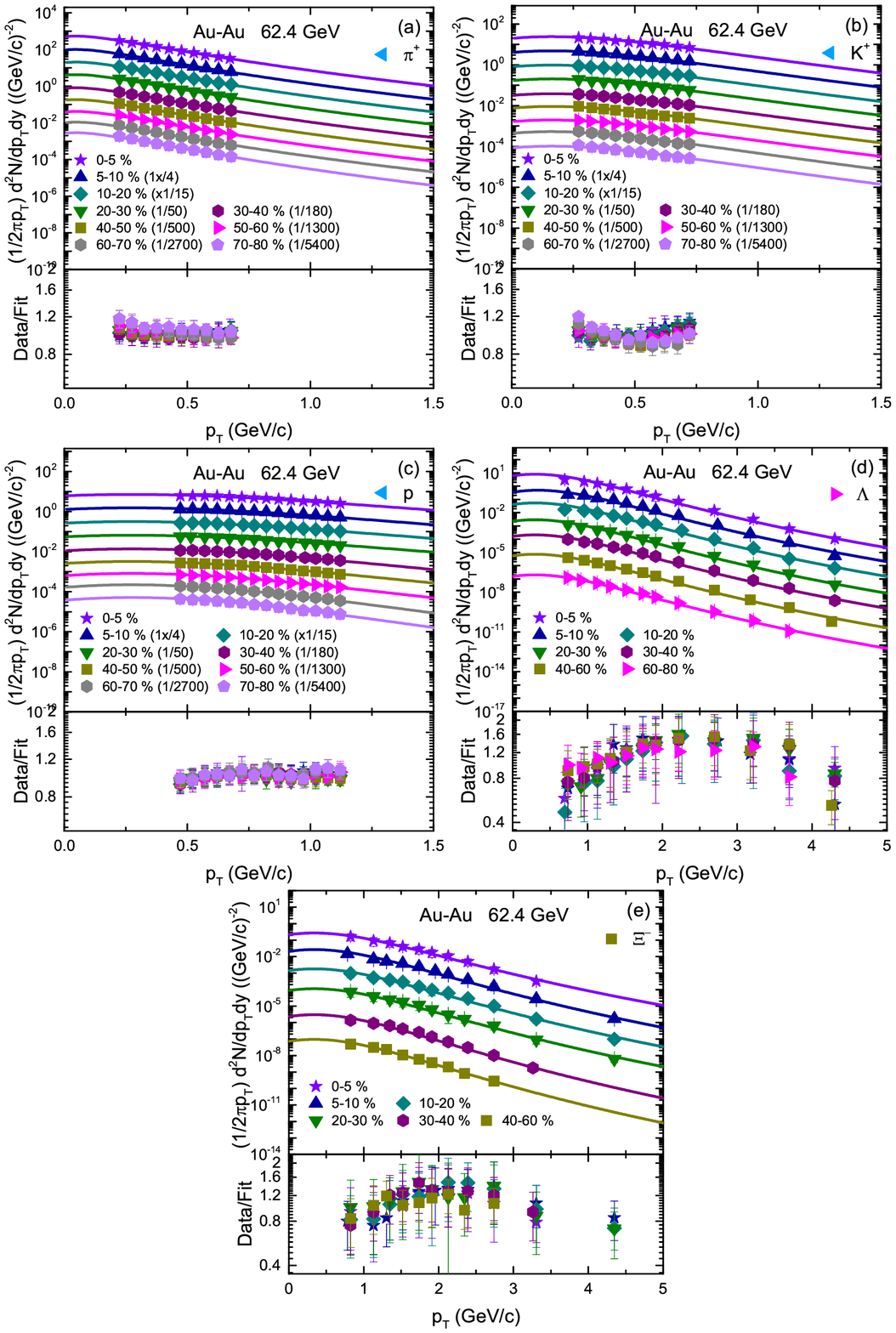}
\end{center}
\justifying\noindent {Fig. 2. The transverse momentum spectra of $\pi^+$, $K^+$,
$p$, $\Lambda$ and $\Xi^-$  in Gold-Gold (Au-Au) collisions at 62.4 GeV in various centrality classes. The experimental data of STAR Collaboration at RHIC for $\pi^+$, $K^+$ and $p$ are taken from ref. \cite{3c} at $|y|<$0.1 rapidity, while for $\Lambda$ and $\Xi$ is taken from ref. \cite{4c} at $|\eta|<$1.8}
\end{figure*}

\begin{figure*}[htbp]
\begin{center}
\includegraphics[width=15.0cm]{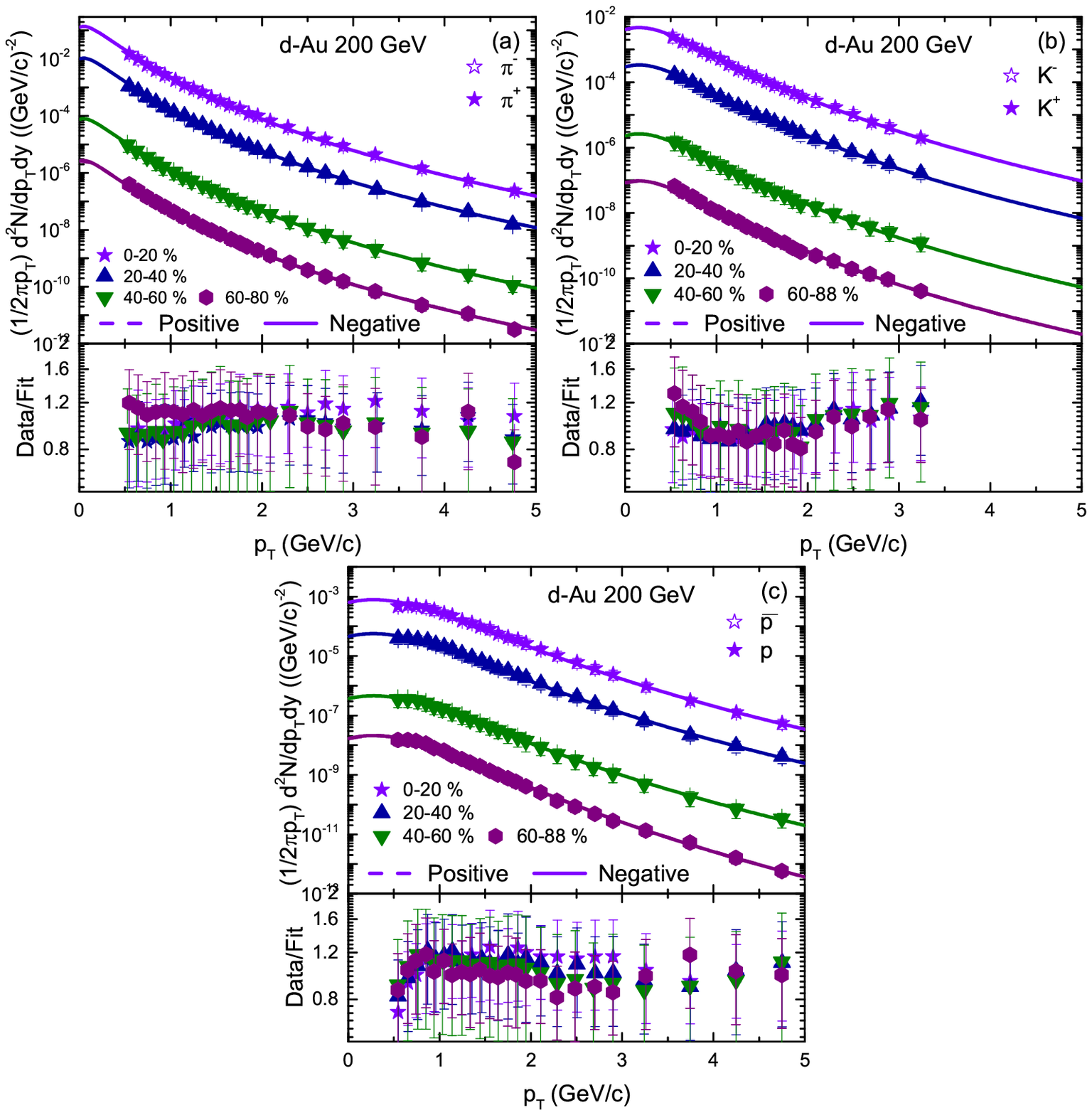}
\end{center}
\justifying\noindent {Fig. 3. The transverse momentum spectra of $\pi^+$, $\pi^-$, $+$, $K^-$, $p$ and
$\bar p$ in Deuteron-Gold (d-Au) collisions at 200 GeV in various centrality intervals and the experimental data of PHENIX Collaboration at RHIC are taken from ref. \cite{5c} at $|\eta|<$0.35.}
\end{figure*}

\begin{table*}[htb!]\vspace{0.2cm}
\vspace{0.25cm} \justifying\noindent {\small Table 1. Values of
free parameters ($T$ and $q$), normalization constant
($\sigma_0$), $\chi^2$, and ndof corresponding to the curves in
Figure 1 for $pp$ collisions at 200 GeV, where the spectrum form
and particle mass are given together. In the last column, ``--"
for ndof denotes the case which has the number of data points
being less than or equal to 0 and the curve is only to guide the
eyes. After rounding, the value of $\chi^2$ is taken to be an
integer. If the integer is 0, we keep one decimal; If the one
decimal is 0.0, we keep two decimals, and so on. \vspace{-0.35cm}

\begin{center}
\newcommand{\tabincell}[2]{\begin{tabular}{@{}#1@{}}#2\end{tabular}}
\begin{tabular}{ccccccccccc}\\ \hline\hline
Collisions       & Centrality       & Particle          & $T_0$ (GeV)   & $\beta_T$ (c) & $V (fm^3)$   & $n$            & $N_0$            & $\chi^2$/ dof \\ \hline
Fig. 1           & 0--10\%          & $\pi^+$      &$0.057\pm0.006$  & $0.309\pm0.012$  & $2200\pm150$  & $10\pm1$     &$55\pm3$             & 0.5/8\\
Cu-Cu            & 10--30\%         & --           &$0.052\pm0.005$  & $0.309\pm0.009$  & $2100\pm130$  & $10\pm1.2$   &$33\pm2$             & 0.6/8\\
200 GeV          & 30--50\%         &--            &$0.049\pm0.005$  & $0.309\pm0.010$  & $2000\pm130$  & $10\pm1$     &$14\pm1$             & 1.5/8\\
                 & 50--70\%         & --           &$0.045\pm0.004$  & $0.309\pm0.010$  & $1874\pm114$  & $10\pm1.2$   &$5.8\pm0.8$          & 0.5/8\\
\cline{2-8}
                 & 0--10\%          & $\pi^-$      &$0.057\pm0.006$  & $0.309\pm0.012$  & $2200\pm150$  & $10\pm1$     &$55\pm3$             & 0.5/8\\
                 & 10--30\%         & --           &$0.052\pm0.005$  & $0.309\pm0.009$  & $2100\pm130$  & $10\pm1.2$   &$33\pm2$             & 0.6/8\\
                 & 30--50\%         &--            &$0.049\pm0.005$  & $0.309\pm0.010$  & $2000\pm130$  & $10\pm1$     &$14\pm1$             & 1.5/8\\
                 & 50--70\%         & --           &$0.045\pm0.004$  & $0.309\pm0.010$  & $1874\pm114$  & $10\pm1.2$   &$5.8\pm0.8$          & 0.5/8\\
\cline{2-8}
                 & 0--10\%          & $K^+$        &$0.076\pm0.004$  & $0.289\pm0.009$  & $2050\pm110$  & $11\pm1.4$   &$0.036\pm0.005$      & 2/7\\
                 & 10--30\%         & --           &$0.072\pm0.004$  & $0.289\pm0.008$  & $1938\pm106$  & $11\pm1.3$   &$0.009\pm0.0005$     & 4/7\\
                 & 30--50\%         &--            &$0.064\pm0.006$  & $0.289\pm0.009$  & $1841\pm116$  & $11\pm1.2$   &$3.8\times10^{-3}\pm5\times10^{-4}$  & 3/7\\
                 & 50--70\%         & --           &$0.060\pm0.004$  & $0.289\pm0.010$  & $1741\pm111$  & $11\pm1.2$   &$3.5\times10^{-3}\pm5\times10^{-4}$  & 8/7\\
\cline{2-8}
                 & 0--10\%          & $K^-$        &$0.076\pm0.004$  & $0.289\pm0.009$  & $2050\pm110$  & $11\pm1.4$   &$0.036\pm0.005$      & 2/7\\
                 & 10--30\%         & --           &$0.072\pm0.004$  & $0.289\pm0.008$  & $1938\pm106$  & $11\pm1.3$   &$0.009\pm0.0005$     & 4/7\\
                 & 30--50\%         &--            &$0.064\pm0.006$  & $0.289\pm0.009$  & $1841\pm116$  & $11\pm1.2$   &$3.8\times10^{-3}\pm5\times10^{-4}$  & 3/7\\
                 & 50--70\%         & --           &$0.060\pm0.004$  & $0.289\pm0.010$  & $1741\pm111$  & $11\pm1.2$   &$3.5\times10^{-3}\pm5\times10^{-4}$  & 8/7\\
\cline{2-8}
                 & 0--10\%       & $p$             &$0.058\pm0.006$  & $0.267\pm0.012$  & $1857\pm123$  & $15\pm1.7$   &$8\times10^{-5}\pm6\times10^{-6}$   & 17/8\\
                 & 10--30\%         & --           &$0.053\pm0.005$  & $0.267\pm0.010$  & $1730\pm124$  & $15\pm2$     &$5\times10^{-5}\pm4\times10^{-6}$   & 22/8\\
                 & 30--50\%         &--            &$0.050\pm0.004$  & $0.267\pm0.010$  & $1600\pm126$  & $15\pm2$     &$2\times10^{-5}\pm4\times10^{-6}$   & 17/8\\
                 & 50--70\%         & --           &$0.046\pm0.005$  & $0.267\pm0.010$  & $1500\pm130$  & $15\pm2.1$   &$7\times10^{-6}\pm4\times10^{-7}$   & 43/8\\
\cline{2-8}
                 & 0--10\%       & $\bar p$        &$0.058\pm0.006$  & $0.267\pm0.012$  & $1857\pm123$  & $15\pm1.7$   &$8\times10^{-5}\pm6\times10^{-6}$   & 17/8\\
                 & 10--30\%         & --           &$0.053\pm0.005$  & $0.267\pm0.010$  & $1730\pm124$  & $15\pm2$     &$5\times10^{-5}\pm4\times10^{-6}$   & 22/8\\
                 & 30--50\%         &--            &$0.050\pm0.004$  & $0.267\pm0.010$  & $1600\pm126$  & $15\pm2$     &$2\times10^{-5}\pm4\times10^{-6}$   & 17/8\\
                 & 50--70\%         & --           &$0.046\pm0.005$  & $0.267\pm0.010$  & $1500\pm130$  & $15\pm2.1$   &$7\times10^{-6}\pm4\times10^{-7}$   & 43/8\\
\cline{2-8}
                 & 0--10\%       & $\Lambda$       &$0.076\pm0.005$  & $0.254\pm0.010$  & $1713\pm108$  & $15.4\pm1.5$ &$2.4\pm0.3$                         & 22/14\\
                 & 10--20\%         & --           &$0.072\pm0.004$  & $0.254\pm0.010$  & $1610\pm104$  & $15.4\pm1.8$ &$0.165\pm0.03$                      & 29/14\\
                 & 20--30\%         &--            &$0.068\pm0.004$  & $0.254\pm0.011$  & $1500\pm111$  & $15.4\pm2.1$ &$0.012\pm0.004$                     & 20/14\\
                 & 30--40\%         & --           &$0.064\pm0.004$  & $0.254\pm0.011$  & $1410\pm120$  & $15.4\pm2.2$ &$8\times10^{-4}\pm5\times10^{-5}$   & 31/14\\
                 & 40--60\%         & --           &$0.060\pm0.006$  & $0.254\pm0.010$  & $1300\pm120$  & $15.4\pm2.2$ &$4.3\times10^{-5}\pm4\times10^{-6}$ & 14/14\\
\cline{2-8}
                 & 0--10\%       & $\Xi$           &$0.076\pm0.004$  & $0.244\pm0.009$  & $1500\pm115$  & $15.5\pm1.5$ &$0.24\pm0.04$                       & 25/6\\
                 & 10--20\%         & --           &$0.072\pm0.004$  & $0.244\pm0.010$  & $1420\pm104$  & $15.5\pm2.2$ &$0.016\pm0.005$                     & 10/6\\
                 & 20--30\%         &--            &$0.069\pm0.004$  & $0.244\pm0.012$  & $1290\pm126$  & $15.5\pm2.2$ &$0.001\pm0.0004$                    & 33/6\\
                 & 30--40\%         & --           &$0.065\pm0.005$  & $0.244\pm0.011$  & $1200\pm100$  & $15.5\pm2.2$ &$8\times10^{-5}\pm5\times10^{-6}$   & 17/6\\
                 & 40--60\%         & --           &$0.061\pm0.004$  & $0.244\pm0.011$  & $1120\pm100$  & $15.5\pm2.2$ &$4\times10^{-6}\pm6\times10^{-7}$   & 15/6\\
\cline{2-8}
Fig. 2           & 0--5\%           & $\pi^+$      &$0.060\pm0.005$  & $0.325\pm0.012$  & $2357\pm170$  & $10.4\pm1.5$ &$270\pm32$                          & 0.2/5\\
Au-Au            & 5--10\%          & --           &$0.056\pm0.006$  & $0.325\pm0.010$  & $2268\pm140$  & $10.4\pm1.5$ &$210\pm18$                         & 0.8/5\\
62.4 GeV         & 10--20\%         &--            &$0.052\pm0.004$  & $0.325\pm0.010$  & $2150\pm130$  & $10.4\pm1$   &$170\pm10$                         & 0.5/5\\
                 & 20--30\%         & --           &$0.048\pm0.005$  & $0.325\pm0.010$  & $2035\pm120$  & $10.4\pm2$   &$146\pm8$                          & 0.5/5\\
                 & 30--40\%         &--            &$0.045\pm0.006$  & $0.325\pm0.011$  & $1955\pm110$  & $10.4\pm1.5$ &$106\pm10$                          & 1.3/5\\
                 & 40--50\%         & --           &$0.042\pm0.005$  & $0.325\pm0.011$  & $1775\pm132$  & $10.4\pm1.2$ &$62\pm8$                            & 1.5/5\\
                 & 50--60\%         &--            &$0.040\pm0.004$  & $0.325\pm0.009$  & $1702\pm108$  & $10.4\pm1.8$ &$38\pm4$                            & 2.5/5\\
                 & 60--70\%         & --           &$0.037\pm0.004$  & $0.325\pm0.011$  & $1620\pm110$  & $10.4\pm1.6$ &$21.7\pm3.5$                        & 5.5/5\\
                 & 70--80\%         & --           &$0.034\pm0.004$  & $0.325\pm0.011$  & $1530\pm88$   & $10.8\pm1.6$ &$11.6\pm2$                          & 2.5/5\\
\cline{2-8}
                 & 0--5\%           & $K^+$        &$0.076\pm0.006$  & $0.312\pm0.010$  & $2100\pm120$  & $11\pm1.5$   &$26.4\pm3.8$                        & 4/5\\
                 & 5--10\%          & --           &$0.073\pm0.006$  & $0.312\pm0.010$  & $2000\pm100$  & $11\pm1.3$   &$22.4\pm1.3$                        & 3/5\\
                 & 10--20\%         &--            &$0.070\pm0.005$  & $0.312\pm0.011$  & $1900\pm100$  & $11\pm1.4$   &$17.7\pm2$                          & 5/5\\
                 & 20--30\%         & --           &$0.067\pm0.005$  & $0.312\pm0.011$  & $1813\pm104$  & $11\pm1.2$   &$13\pm2$                            & 4/5\\
                 & 30--40\%         &--            &$0.064\pm0.005$  & $0.312\pm0.010$  & $1730\pm108$  & $11\pm1.4$   &$9.2\pm1.4$                         & 4/5\\
                 & 40--50\%         & --           &$0.061\pm0.004$  & $0.312\pm0.010$  & $1650\pm102$  & $11\pm1.3$   &$6.4\pm1.2$                         & 10/5\\
                 & 50--60\%         &--            &$0.057\pm0.004$  & $0.312\pm0.010$  & $1560\pm85$   & $11\pm1.2$   &$3.8\pm0.4$                         & 2/5\\
                 & 60--70\%         & --           &$0.054\pm0.005$  & $0.312\pm0.009$  & $1481\pm70$   & $11\pm1.1$   &$2.2\pm0.4$                         & 10/5\\
                 & 70--80\%         & --           &$0.050\pm0.006$  & $0.312\pm0.009$  & $1400\pm80$   & $11\pm1.2$   &$0.9\pm0.03$                        & 13/5\\
\hline
\end{tabular}%
\end{center}}
\end{table*}

\begin{table*}[htb!]\vspace{0.2cm}
\vspace{0.25cm} \justifying\noindent {\small Table 1. Continue. \vspace{-0.35cm}

\begin{center}
\newcommand{\tabincell}[2]{\begin{tabular}{@{}#1@{}}#2\end{tabular}}
\begin{tabular} {cccccccccccc}\\ \hline\hline
Collisions       & Centrality       & Particle          & $T_0$ (GeV)   & $\beta_T$ (c) & $V (fm^3)$   & $n$            & $N_0$            & $\chi^2$/ dof \\ \hline
                 & 0--5\%           & $p$          &$0.060\pm0.005$  & $0.280\pm0.010$  & $1930\pm108$  & $8.4\pm1$    &$7\pm0.8$                           & 9/9\\
                 & 5--10\%          & --           &$0.056\pm0.005$  & $0.280\pm0.011$  & $1843\pm110$  & $9\pm0.6$    &$6.1\pm0.6$                         & 2/9\\
                 & 10--20\%         &--            &$0.053\pm0.005$  & $0.280\pm0.009$  & $1739\pm120$  & $9\pm0.4$    &$5\pm0.5$                           & 1/9\\
                 & 20--30\%         & --           &$0.049\pm0.004$  & $0.280\pm0.010$  & $1627\pm109$  & $9\pm1$      &$3.7\pm0.4$                         & 3/9\\
                 & 30--40\%         &--            &$0.045\pm0.006$  & $0.280\pm0.012$  & $1500\pm100$  & $11\pm1$     &$2.8\pm0.5$                         & 2/9\\
                 & 40--50\%         & --           &$0.042\pm0.005$  & $0.280\pm0.012$  & $1385\pm100$  & $11.5\pm1.5$ &$1.95\pm0.3$                        & 2/9\\
                 & 50--60\%         &--            &$0.040\pm0.005$  & $0.280\pm0.011$  & $1300\pm93$   & $13\pm1.6$   &$1.3\pm0.04$                        & 1/9\\
                 & 60--70\%         & --           &$0.038\pm0.004$  & $0.280\pm0.010$  & $1212\pm80$   & $15\pm1.4$   &$0.745\pm0.04$                      & 3/9\\
                 & 70--80\%         & --           &$0.035\pm0.005$  & $0.280\pm0.010$  & $1100\pm60$   & $16\pm1.5$   &$0.37\pm0.03$                       & 4/9\\
\cline{2-8}
                 & 0--5\%           & $\Lambda$    &$0.076\pm0.006$  & $0.260\pm0.012$  & $1800\pm126$  & $15\pm1.5$   &$7.6\pm0.5$                         & 5/7\\
                 & 5--10\%          & --           &$0.073\pm0.006$  & $0.260\pm0.009$  & $1700\pm100$  & $15\pm1.6$   &$0.61\pm0.06$                       & 11/7\\
                 & 10--20\%         &--            &$0.070\pm0.006$  & $0.260\pm0.011$  & $1610\pm100$  & $15\pm1.5$   &$0.058\pm0.005$                     & 8/7\\
                 & 20--30\%         & --           &$0.067\pm0.005$  & $0.260\pm0.013$  & $1523\pm96$   & $15\pm1.7$   &$0.0033\pm0.0004$                   & 8/7\\
                 & 30--40\%         &--            &$0.064\pm0.005$  & $0.260\pm0.010$  & $1450\pm70$   & $15\pm1.4$   &$2.5\times10^{-4}\pm4\times10^{-5}$ & 3/7\\
                 & 40--60\%         & --           &$0.060\pm0.004$  & $0.260\pm0.011$  & $1329\pm80$   & $15\pm1.6$   &$9.4\times10^{-6}\pm5\times10^{-7}$  & 16/7\\
                 & 60--80\%         &--            &$0.055\pm0.004$  & $0.260\pm0.010$  & $1200\pm85$   & $15\pm1.6$   &$2.9\times10^{-7}\pm4\times10^{-8}$ & 3/7\\
\cline{2-8}
                 & 0--5\%           & $\Xi^-$      &$0.076\pm0.006$  & $0.249\pm0.012$  & $1680\pm100$  & $13\pm1.2$   &$0.33\pm0.04$                       & 2/5\\
                 & 5--10\%          & --           &$0.073\pm0.006$  & $0.249\pm0.009$  & $1600\pm100$  & $14\pm1.2$   &$0.033\pm0.004$                     & 4/5\\
                 & 10--20\%         &--            &$0.070\pm0.006$  & $0.249\pm0.011$  & $1500\pm80$   & $14\pm1.3$   &$0.0024\pm0.0005$                   & 4/5\\
                 & 20--30\%         & --           &$0.067\pm0.005$  & $0.249\pm0.013$  & $1400\pm70$   & $14\pm1.5$   &$1.6\times10^{-4}\pm5\times10^{-5}$ & 1.5/5\\
                 & 30--40\%         &--            &$0.064\pm0.005$  & $0.249\pm0.010$  & $1320\pm70$   & $15\pm1.5$   &$4.4\times10^{-6}\pm5\times10^{-7}$ & 3/5\\
                 & 40--60\%         & --           &$0.060\pm0.004$  & $0.249\pm0.011$  & $1200\pm80$   & $15\pm1.5$   &$1.4\times10^{-7}\pm5\times10^{-8}$ & 1.5/5\\
\cline{2-8}
Fig. 3           & 0--20\%          & $\pi^+$      &$0.058\pm0.004$  & $0.315\pm0.009$  & $2250\pm100$  & $9.8\pm1.3$  &$0.075\pm0.005$                     & 2/19\\
d-Au             & 20--40\%         & --           &$0.055\pm0.004$  & $0.315\pm0.009$  & $2170\pm104$  & $9.8\pm1.3$  &$0.006\pm0.0004$                    & 2/19\\
 200 GeV         & 40--60\%         & --           &$0.048\pm0.005$  & $0.315\pm0.011$  & $1954\pm110$  & $9.8\pm1.2$  &$5\times10^{-5}\pm5\times10^{-6}$   & 0.6/19\\
                 & 60--80\%         &--            &$0.044\pm0.004$  & $0.315\pm0.009$  & $1843\pm123$  & $9.8\pm1.4$  &$1.8\times10^{-6}\pm4\times10^{-7}$ & 3/19\\
\cline{2-8}
                 & 0--20\%           & $\pi^-$      &$0.058\pm0.004$  & $0.315\pm0.009$  & $2250\pm100$  & $9.8\pm1.3$  &$0.075\pm0.005$                     & 2/19\\
                 & 20--40\%          & --           &$0.055\pm0.004$  & $0.315\pm0.009$  & $2170\pm104$  & $9.8\pm1.3$  &$0.006\pm0.0004$                    & 2/19\\
                 & 40--60\%         & --           &$0.048\pm0.005$  & $0.315\pm0.011$  & $1954\pm110$  & $9.8\pm1.2$  &$5\times10^{-5}\pm5\times10^{-6}$   & 0.6/19\\
                 & 60--80\%         &--            &$0.044\pm0.004$  & $0.315\pm0.009$  & $1843\pm123$  & $9.8\pm1.4$  &$1.8\times10^{-6}\pm4\times10^{-7}$ & 3/19\\
\cline{2-8}
                 & 0--20\%           & $K^+$        &$0.078\pm0.005$  & $0.296\pm0.010$  & $2119\pm110$  & $9.4\pm1.2$  &$0.0055\pm0.0005$                   & 1.4/14\\
                 & 20--40\%          & --           &$0.074\pm0.006$  & $0.296\pm0.011$  & $2041\pm121$  & $9.4\pm1.4$  &$4.1\times10^{-4}\pm5\times10^{-5}$ & 1/14\\
                 & 40--60\%         & --           &$0.067\pm0.006$  & $0.296\pm0.013$  & $1817\pm116$  & $9.4\pm1.4$  &$3.6\times10^{-6}\pm5\times10^{-7}$ & 1/14\\
                 & 60--80\%         &--            &$0.062\pm0.005$  & $0.296\pm0.011$  & $1703\pm120$  & $9.8\pm1.1$  &$1.4\times10^{-7}\pm6\times10^{-8}$ & 3/14\\
\cline{2-8}
                 & 0--20\%           & $K^-$        &$0.078\pm0.005$  & $0.296\pm0.010$  & $2119\pm110$  & $9.4\pm1.2$  &$0.0055\pm0.0005$                   & 1.4/14\\
                 & 20--40\%          & --           &$0.074\pm0.006$  & $0.296\pm0.011$  & $2041\pm121$  & $9.4\pm1.4$  &$4.1\times10^{-4}\pm5\times10^{-5}$ & 1/14\\
                 & 40--60\%         & --           &$0.067\pm0.006$  & $0.296\pm0.013$  & $1817\pm116$  & $9.4\pm1.4$  &$3.6\times10^{-6}\pm5\times10^{-7}$ & 1/14\\
                 & 60--80\%         &--            &$0.062\pm0.005$  & $0.296\pm0.011$  & $1703\pm120$  & $9.8\pm1.1$  &$1.4\times10^{-7}\pm6\times10^{-8}$ & 3/14\\
\cline{2-8}
                 & 0--20\%           & $p$          &$0.061\pm0.006$  & $0.283\pm0.008$  & $1920\pm109$  & $11\pm1.5$   &$7.3\times10^{-4}\pm6\times10^{-5}$ & 4.5/19\\
                 & 20--40\%          & --           &$0.057\pm0.004$  & $0.283\pm0.010$  & $1834\pm101$  & $11\pm1.5$   &$5.5\times10^{-5}\pm5\times10^{-6}$ & 2/19\\
                 & 40--60\%         & --           &$0.048\pm0.005$  & $0.283\pm0.011$  & $1600\pm106$  & $11\pm1.2$   &$5.1\times10^{-7}\pm4\times10^{-8}$ & 1/19\\
                 & 60--80\%         &--            &$0.045\pm0.004$  & $0.283\pm0.010$  & $1500\pm102$  & $12\pm1.4$   &$2.4\times10^{-8}\pm5\times10^{-9}$ & 2/19\\
\cline{2-8}
                 & 0--20\%           & $\bar p$     &$0.061\pm0.006$  & $0.283\pm0.008$  & $1920\pm109$  & $11\pm1.5$   &$7.3\times10^{-4}\pm6\times10^{-5}$ & 4.5/19\\
                 & 20--40\%          & --           &$0.057\pm0.004$  & $0.283\pm0.010$  & $1834\pm101$  & $11\pm1.5$   &$5.5\times10^{-5}\pm5\times10^{-6}$ & 2/19\\
                 & 40--60\%         & --           &$0.048\pm0.005$  & $0.283\pm0.011$  & $1600\pm106$  & $11\pm1.2$   &$5.1\times10^{-7}\pm4\times10^{-8}$ & 1/19\\
                 & 60--80\%         &--            &$0.045\pm0.004$  & $0.283\pm0.010$  & $1500\pm102$  & $12\pm1.4$   &$2.4\times10^{-8}\pm5\times10^{-9}$ & 2/19\\
\end{tabular}%
\end{center}}
\end{table*}

\begin{table*}[htb!]\vspace{0.2cm}
\vspace{0.25cm} \justifying\noindent {\small Table 1. Continue. \vspace{-0.35cm}

\begin{center}
\newcommand{\tabincell}[2]{\begin{tabular}{@{}#1@{}}#2\end{tabular}}
\begin{tabular}{ccccccccccc}\\ \hline\hline
Collisions       & Centrality       & Particle          & $T_0$ (GeV)   & $\beta_T$ (c) & $V (fm^3)$   & $n$            & $N_0$            & $\chi^2$/ dof \\ \hline
Fig. 4           & 0--5\%           & $\pi^+$      &$0.108\pm0.005$  & $0.443\pm0.011$  & $3500\pm140$  & $9.8\pm1$    &$2.3\times10^{5}\pm5\times10^{4}$   & 34/36\\
Pb-Pb            & 5--10\%          & --           &$0.104\pm0.004$  & $0.443\pm0.010$  & $3360\pm143$  & $9.8\pm1$    &$1.1\times10^{5}\pm4\times10^{4}$  & 40/36\\
2.76 TeV         & 10--20\%         &              &$0.100\pm0.005$  & $0.443\pm0.010$  & $3237\pm136$  & $9.8\pm1.2$  &$4\times10^{4}\pm6\times10^{3}$     & 21/36\\
                 & 20--30\%         & --           &$0.096\pm0.006$  & $0.443\pm0.012$  & $3100\pm124$  & $9.8\pm1.3$  &$1.4\times10^{4}\pm4\times10^{3}$   & 29/36\\
                 & 30--40\%         &--            &$0.092\pm0.006$  & $0.443\pm0.010$  & $2974\pm109$  & $9.8\pm1$    &$5\times10^{3}\pm5\times10^{2}$     & 40/36\\
                 & 40--50\%         & --           &$0.089\pm0.004$  & $0.443\pm0.012$  & $2854\pm108$  & $9.8\pm1$    &$1.6\times10^{3}\pm3\times10^{2}$   & 35/36\\
                 & 50--60\%        &--            &$0.086\pm0.005$  & $0.443\pm0.010$  & $2719\pm120$  & $9.8\pm1.1$  &$450\pm21$                          & 34/36\\
                 & 60--70\%        &--            &$0.082\pm0.006$  & $0.443\pm0.012$  & $2600\pm132$  & $10.3\pm1.3$ &$150\pm10$                          & 55/36\\
                 & 70--80\%        &--            &$0.078\pm0.004$  & $0.443\pm0.012$  & $2500\pm100$  & $10.3\pm1.2$ &$34\pm4$                            & 71/36\\
                 & 80--90\%        &--            &$0.074\pm0.005$  & $0.443\pm0.010$  & $2300\pm161$  & $10.3\pm1.4$ &$7.5\pm0.4$                         & 73/36\\
\cline{2-8}
                 & 0--5\%           & $\pi^-$      &$0.108\pm0.005$  & $0.443\pm0.011$  & $3500\pm140$  & $9.8\pm1$    &$2.3\times10^{5}\pm5\times10^{4}$   & 34/36\\
                 & 5--10\%          & --           &$0.104\pm0.004$  & $0.443\pm0.010$  & $3360\pm143$  & $9.8\pm1$    &$1.1\times10^{5}\pm4\times10^{4}$  & 40/36\\
                 & 10--20\%         &              &$0.100\pm0.005$  & $0.443\pm0.010$  & $3237\pm136$  & $9.8\pm1.2$  &$4\times10^{4}\pm6\times10^{3}$     & 21/36\\
                 & 20--30\%         & --           &$0.096\pm0.006$  & $0.443\pm0.012$  & $3100\pm124$  & $9.8\pm1.3$  &$1.4\times10^{4}\pm4\times10^{3}$   & 29/36\\
                 & 30--40\%         &--            &$0.092\pm0.006$  & $0.443\pm0.010$  & $2974\pm109$  & $9.8\pm1$    &$5\times10^{3}\pm5\times10^{2}$     & 40/36\\
                 & 40--50\%         & --           &$0.089\pm0.004$  & $0.443\pm0.012$  & $2854\pm108$  & $9.8\pm1$    &$1.6\times10^{3}\pm3\times10^{2}$   & 35/36\\
                 & 50--60\%        &--             &$0.086\pm0.005$  & $0.443\pm0.010$  & $2719\pm120$  & $9.8\pm1.1$  &$450\pm21$                          & 34/36\\
                 & 60--70\%        &--             &$0.082\pm0.006$  & $0.443\pm0.012$  & $2600\pm132$  & $10.3\pm1.3$ &$150\pm10$                          & 55/36\\
                 & 70--80\%        &--             &$0.078\pm0.004$  & $0.443\pm0.012$  & $2500\pm100$  & $10.3\pm1.2$ &$34\pm4$                            & 71/36\\
                 & 80--90\%        &--             &$0.074\pm0.005$  & $0.443\pm0.010$  & $2300\pm161$  & $10.3\pm1.4$ &$7.5\pm0.4$                         & 73/36\\
\cline{2-8}
                 & 0--5\%           & $K^+$        &$0.129\pm0.004$  & $0.421\pm0.009$  & $3319\pm138$  & $9.3\pm1.2$  &$1.9\times10^{4}\pm4\times10^{3}$   & 40/31\\
                 & 5--10\%          & --           &$0.126\pm0.004$  & $0.421\pm0.009$  & $3319\pm138$  & $9.3\pm1.2$  &$8.6\times10^{3}\pm5\times10^{2}$   & 24/31\\
                 & 10--20\%         &              &$0.122\pm0.005$  & $0.421\pm0.011$  & $3200\pm180$  & $9.3\pm1.3$  &$3.3\times10^{3}\pm5\times10^{2}$   & 25/31\\
                 & 20--30\%         & --           &$0.119\pm0.006$  & $0.421\pm0.010$  & $3000\pm181$  & $9.3\pm1.3$  &$1.2\times10^{3}\pm4\times10^{2}$   & 17/31\\
                 & 30--40\%         &--            &$0.115\pm0.005$  & $0.421\pm0.012$  & $2875\pm130$  & $9.3\pm1.3$  &$400\pm24$                          & 10/31\\
                 & 40--50\%         & --           &$0.110\pm0.006$  & $0.421\pm0.011$  & $2732\pm131$  & $9.3\pm1.4$  &$130\pm10$                          & 16/31\\
                 & 50--60\%        &--             &$0.107\pm0.005$  & $0.421\pm0.010$  & $2600\pm126$  & $9.3\pm1.4$  &$36\pm5$                            & 20/31\\
                 & 60--70\%        &--             &$0.102\pm0.004$  & $0.421\pm0.012$  & $2450\pm142$  & $9.5\pm1.3$  &$10\pm0.4$                          & 12/31\\
                 & 70--80\%        &--             &$0.098\pm0.005$  & $0.421\pm0.010$  & $2350\pm126$  & $9.8\pm1.3$  &$2.5\pm0.3$                         & 65/31\\
                 & 80--90\%        &--             &$0.094\pm0.005$  & $0.421\pm0.011$  & $2220\pm141$  & $10\pm1.4$   &$0.5\pm0.03$                        & 35/31\\
\cline{2-8}
                 & 0--5\%           & $K^-$        &$0.129\pm0.004$  & $0.421\pm0.009$  & $3319\pm138$  & $9.3\pm1.2$  &$1.9\times10^{4}\pm4\times10^{3}$   & 40/31\\
                 & 5--10\%          & --           &$0.126\pm0.004$  & $0.421\pm0.009$  & $3319\pm138$  & $9.3\pm1.2$  &$8.6\times10^{3}\pm5\times10^{2}$   & 24/31\\
                 & 10--20\%         &              &$0.122\pm0.005$  & $0.421\pm0.011$  & $3200\pm180$  & $9.3\pm1.3$  &$3.3\times10^{3}\pm5\times10^{2}$   & 25/31\\
                 & 20--30\%         & --           &$0.119\pm0.006$  & $0.421\pm0.010$  & $3000\pm181$  & $9.3\pm1.3$  &$1.2\times10^{3}\pm4\times10^{2}$   & 17/31\\
                 & 30--40\%         &--            &$0.115\pm0.005$  & $0.421\pm0.012$  & $2875\pm130$  & $9.3\pm1.3$  &$400\pm24$                          & 10/31\\
                 & 40--50\%         & --           &$0.110\pm0.006$  & $0.421\pm0.011$  & $2732\pm131$  & $9.3\pm1.4$  &$130\pm10$                          & 16/31\\
                 & 50--60\%        &--             &$0.107\pm0.005$  & $0.421\pm0.010$  & $2600\pm126$  & $9.3\pm1.4$  &$36\pm5$                            & 20/31\\
                 & 60--70\%        &--             &$0.102\pm0.004$  & $0.421\pm0.012$  & $2450\pm142$  & $9.5\pm1.3$  &$10\pm0.4$                          & 12/31\\
                 & 70--80\%        &--             &$0.098\pm0.005$  & $0.421\pm0.010$  & $2350\pm126$  & $9.8\pm1.3$  &$2.5\pm0.3$                         & 65/31\\
                 & 80--90\%        &--             &$0.094\pm0.005$  & $0.421\pm0.011$  & $2220\pm141$  & $10\pm1.4$   &$0.5\pm0.03$                        & 35/31\\
\cline{2-8}
                 & 0--5\%          & $p$           &$0.108\pm0.005$  & $0.403\pm0.010$  & $3100\pm140$  & $9.8\pm1$    &$2.3\times10^{5}\pm5\times10^{4}$   & 34/32\\
                 & 5--10\%          & --           &$0.104\pm0.006$  & $0.403\pm0.011$  & $2900\pm142$  & $9\pm1.2$    &$1000\pm100$                        & 151/32\\
                 & 10--20\%         &              &$0.101\pm0.004$  & $0.403\pm0.010$  & $2768\pm120$  & $9\pm1.2$    &$400\pm20$                          & 151/32\\
                 & 20--30\%         & --           &$0.097\pm0.005$  & $0.403\pm0.010$  & $2550\pm136$  & $9\pm1.2$    &$150\pm10$                          & 148/32\\
                 & 30--40\%         &--            &$0.094\pm0.005$  & $0.403\pm0.010$  & $2400\pm130$  & $9\pm1.1$    &$47\pm4$                            & 143/32\\
\hline
\end{tabular}%
\end{center}}
\end{table*}

\begin{table*}[htb!]\vspace{0.2cm}
\vspace{0.25cm} \justifying\noindent {\small Table 1. Continue. \vspace{-0.35cm}

\begin{center}
\newcommand{\tabincell}[2]{\begin{tabular}{@{}#1@{}}#2\end{tabular}}
\begin{tabular}{ccccccccccc}\\ \hline\hline
Collisions       & Centrality       & Particle          & $T_0$ (GeV)   & $\beta_T$ (c) & $V (fm^3)$   & $n$            & $N_0$            & $\chi^2$/ dof \\ \hline
& 40--50\%         & --           &$0.090\pm0.005$  & $0.403\pm0.011$  & $2300\pm140$  & $9.8\pm1.3$  &$17\pm2$                            & 43/32\\
                 & 50--60\%        &--             &$0.086\pm0.004$  & $0.403\pm0.012$  & $2200\pm100$  & $10.3\pm1.1$ &$5.5\pm0.5$                         & 17/32\\
                 & 60--70\%        &--             &$0.083\pm0.005$  & $0.403\pm0.012$  & $2050\pm110$  & $10.3\pm1.1$ &$1.6\pm0.3$                         & 35/32\\
                 & 70--80\%        &--             &$0.080\pm0.006$  & $0.403\pm0.012$  & $1910\pm120$  & $10.7\pm1.4$ &$0.38\pm0.02$                       & 54/32\\
                 & 80--90\%        &--             &$0.077\pm0.004$  & $0.403\pm0.010$  & $1800\pm100$  & $11.7\pm1.2$ &$0.08\pm0.003$                      & 61/32\\
\cline{2-8}
                 & 0--5\%          & $\bar p$      &$0.108\pm0.005$  & $0.403\pm0.010$  & $3100\pm140$  & $9.8\pm1$    &$2.3\times10^{5}\pm5\times10^{4}$   & 34/32\\
                 & 5--10\%          & --           &$0.104\pm0.006$  & $0.403\pm0.011$  & $2900\pm142$  & $9\pm1.2$    &$1000\pm100$                        & 151/32\\
                 & 10--20\%         &              &$0.101\pm0.004$  & $0.403\pm0.010$  & $2768\pm120$  & $9\pm1.2$    &$400\pm20$                          & 151/32\\
                 & 20--30\%         & --           &$0.097\pm0.005$  & $0.403\pm0.010$  & $2550\pm136$  & $9\pm1.2$    &$150\pm10$                          & 148/32\\
                 & 30--40\%         &--            &$0.094\pm0.005$  & $0.403\pm0.010$  & $2400\pm130$  & $9\pm1.1$    &$47\pm4$                            & 143/32\\
                 & 40--50\%         & --           &$0.090\pm0.005$  & $0.403\pm0.011$  & $2300\pm140$  & $9.8\pm1.3$  &$17\pm2$                            & 43/32\\
                 & 50--60\%        &--             &$0.086\pm0.004$  & $0.403\pm0.012$  & $2200\pm100$  & $10.3\pm1.1$ &$5.5\pm0.5$                         & 17/32\\
                 & 60--70\%        &--             &$0.083\pm0.005$  & $0.403\pm0.012$  & $2050\pm110$  & $10.3\pm1.1$ &$1.6\pm0.3$                         & 35/32\\
                 & 70--80\%        &--             &$0.080\pm0.006$  & $0.403\pm0.012$  & $1910\pm120$  & $10.7\pm1.4$ &$0.38\pm0.02$                       & 54/32\\
                 & 80--90\%        &--             &$0.077\pm0.004$  & $0.403\pm0.010$  & $1800\pm100$  & $11.7\pm1.2$ &$0.08\pm0.003$                      & 61/32\\
\cline{2-8}
                 & 0--10\%         & $\Lambda$     &$0.128\pm0.005$  & $0.381\pm0.010$  & $2900\pm150$  & $7\pm0.7$    &$17\pm2$                            & 130/14\\
                 & 10--20\%        &--             &$0.126\pm0.006$  & $0.381\pm0.010$  & $2750\pm100$  & $7\pm1$      &$13\pm0.4$                          & 74/14\\
                 & 20--40\%        &--             &$0.120\pm0.005$  & $0.381\pm0.011$  & $2623\pm110$  & $7.3\pm1.1$  &$7.9\pm0.2$                         & 105/14\\
                 & 40--60\%        &--             &$0.111\pm0.006$  & $0.381\pm0.012$  & $2500\pm130$  & $9.4\pm1.2$  &$3.5\pm0.2$                         & 22/14\\
                 & 60--80\%        &--             &$0.105\pm0.005$  & $0.381\pm0.012$  & $2400\pm152$  & $10.5\pm1.3$ &$1\pm0.02$                          & 20/14\\
\cline{2-8}
                 & 0--10\%         & $\Xi$         &$0.128\pm0.006$  & $0.287\pm0.012$  & $2600\pm180$  & $7\pm0.6$    &$5.5\pm0.3$                         & 39/7\\
                 & 10--20\%        &--             &$0.125\pm0.004$  & $0.287\pm0.012$  & $2477\pm150$  & $7\pm0.5$    &$4.2\pm0.4$                         & 25/7\\
                 & 20--40\%        &--             &$0.120\pm0.004$  & $0.287\pm0.010$  & $2356\pm144$  & $7.6\pm0.4$  &$2.7\pm0.2$                         & 13/7\\
                 & 40--60\%        &--             &$0.112\pm0.005$  & $0.287\pm0.011$  & $2207\pm108$  & $8\pm0.4$    &$1\pm0.2$                           & 32/7\\
                 & 60--80\%        &--             &$0.105\pm0.004$  & $0.287\pm0.010$  & $2100\pm120$  & $10\pm0.8$   &$0.3\pm0.03$                        & 57/7\\
\cline{2-8}
Fig. 5           & 0--5\%          & $\pi^++\pi^-$ &$0.100\pm0.006$  & $0.430\pm0.010$  & $3300\pm150$  & $9\pm1$      &$1700\pm113$                        & 67/36\\
p-Pb             & 5--10\%          & --           &$0.096\pm0.005$  & $0.430\pm0.011$  & $3200\pm136$  & $9\pm1$      &$700\pm100$                         & 51/36\\
5.02 TeV         & 10--20\%         &--            &$0.092\pm0.004$  & $0.430\pm0.012$  & $3100\pm124$  & $9\pm1.2$    &$300\pm13$                         & 44/36\\
                 & 20--40\%         & --           &$0.088\pm0.005$  & $0.430\pm0.011$  & $2942\pm156$  & $9\pm1.1$    &$130\pm16$                         & 49/36\\
                 & 40--60\%         &--            &$0.083\pm0.006$  & $0.430\pm0.009$  & $2800\pm140$  & $9\pm1.2$    &$37\pm7$                            & 42/36\\
                 & 60--80\%         & --           &$0.079\pm0.005$  & $0.430\pm0.011$  & $2647\pm137$  & $9.7\pm1.3$  &$17\pm3$                            & 82/36\\
                 & 80--100\%        &--            &$0.075\pm0.005$  & $0.430\pm0.011$  & $2500\pm141$  & $10\pm1.2$   &$3.6\pm0.5$                         & 67/36\\
\cline{2-8}
                 & 0--5\%           & $K^++K^-$    &$0.120\pm0.005$  & $0.400\pm0.011$  & $3100\pm145$  & $8\pm0.7$    &$128\pm13$                          & 5/26\\
                 & 5--10\%          & --           &$0.116\pm0.006$  & $0.400\pm0.012$  & $2974\pm130$  & $8\pm0.5$    &$52\pm4$                            & 8/26\\
                 & 10--20\%         &--            &$0.112\pm0.005$  & $0.400\pm0.012$  & $2851\pm140$  & $8.2\pm0.6$  &$23\pm3$                            & 45/26\\
                 & 20--40\%         & --           &$0.108\pm0.005$  & $0.400\pm0.011$  & $2714\pm124$  & $8.2\pm0.5$  &$10\pm0.4$                          & 10/26\\
                 & 40--60\%         &--            &$0.104\pm0.005$  & $0.400\pm0.010$  & $2600\pm108$  & $8.6\pm0.6$  &$5.2\pm0.4$                         & 21/26\\
                 & 60--80\%         & --           &$0.100\pm0.005$  & $0.400\pm0.010$  & $2500\pm100$  & $9\pm0.7$    &$1.2\pm0.04$                        & 45/26\\
                 & 80--100\%        &--            &$0.096\pm0.004$  & $0.400\pm0.011$  & $2400\pm100$  & $10.6\pm1$   &$0.3\pm0.03$                        & 90/26\\
\cline{2-8}
                 & 0--5\%           & $p+\bar p$   &$0.100\pm0.004$  & $0.380\pm0.012$  & $2928\pm140$  & $8.4\pm0.8$  &$20\pm2$                            & 9/34\\
                 & 5--10\%          & --           &$0.096\pm0.005$  & $0.380\pm0.010$  & $2800\pm120$  & $8.4\pm0.6$  &$8.7\pm0.6$                         & 14/34\\
                 & 10--20\%         &--            &$0.092\pm0.005$  & $0.380\pm0.010$  & $2700\pm130$  & $8.4\pm0.6$  &$3.5\pm0.2$                         & 50/34\\
                 & 20--40\%         & --           &$0.088\pm0.006$  & $0.380\pm0.010$  & $2570\pm110$  & $8.9\pm0.8$  &$1.8\pm0.03$                        & 86/34\\
                 & 40--60\%         &--            &$0.084\pm0.005$  & $0.380\pm0.011$  & $2450\pm109$  & $9.5\pm0.5$  &$0.6\pm0.02$                        & 6/34\\
                 & 60--80\%         & --           &$0.080\pm0.005$  & $0.380\pm0.011$  & $2323\pm108$  & $10.5\pm0.6$ &$0.24\pm0.02$                       & 28/34\\
                 & 80--100\%        &--            &$0.078\pm0.005$  & $0.380\pm0.010$  & $2200\pm100$  & $12\pm1$     &$0.059\pm0.003$                     & 91/34\\
\cline{2-8}
                 & 0--5\% & $\Lambda+\bar \Lambda$ &$0.120\pm0.006$  & $0.330\pm0.009$  & $2500\pm180$  & $8.8\pm0.4$  &$15\pm2$                            & 159/13\\
                 & 5--10\%          & --           &$0.115\pm0.006$  & $0.330\pm0.011$  & $2350\pm135$  & $8.8\pm0.3$  &$7\pm0.4$                           & 96/13\\
                 & 10--20\%         &--            &$0.114\pm0.004$  & $0.330\pm0.012$  & $2220\pm146$  & $8.8\pm0.4$  &$3\pm0.2$                           & 110/13\\
                 & 20--40\%         & --           &$0.110\pm0.005$  & $0.330\pm0.012$  & $2220\pm142$  & $9\pm0.35$   &$1.15\pm0.05$                       & 32/13\\
                 & 40--60\%         &--            &$0.105\pm0.004$  & $0.330\pm0.010$  & $2100\pm136$  & $9\pm0.5$    &$0.4\pm0.02$                        & 73/13\\
                 & 60--80\%         & --           &$0.101\pm0.005$  & $0.330\pm0.012$  & $2000\pm150$  & $9.9\pm0.7$  &$0.14\pm0.02$                       & 136/13\\
                 & 80--100\%        &--            &$0.096\pm0.005$  & $0.330\pm0.010$  & $1870\pm108$  & $10.4\pm0.8$ &$0.27\pm0.05$                       & 376/13\\
  \hline
\end{tabular}%
\end{center}}
\end{table*}

\begin{table*}[htb!]\vspace{0.2cm}
\vspace{0.25cm} \justifying\noindent {\small Table 1. Continue. \vspace{-0.35cm}

\begin{center}
\newcommand{\tabincell}[2]{\begin{tabular}{@{}#1@{}}#2\end{tabular}}
\begin{tabular}{ccccccccccc}\\ \hline\hline
Collisions       & Centrality       & Particle          & $T_0$ (GeV)   & $\beta_T$ (c) & $V (fm^3)$   & $n$            & $N_0$            & $\chi^2$/ dof \\ \hline             p-p 200 GeV      & ---             & $\pi^+$       &$0.042\pm0.006$  & $0.256\pm0.008$  & $532\pm30$    & $8\pm0.4$    &$63\pm3$                            & 9/22\\
                 & ---             & $\pi^-$       &$0.042\pm0.006$  & $0.256\pm0.008$  & $532\pm30$    & $8\pm0.4$    &$63\pm3$                            & 9/22\\
                 & ---             & $K^+$         &$0.050\pm0.004$  & $0.230\pm0.009$  & $434\pm30$    & $8\pm0.4$    &$10\pm0.3$                          & 8/11\\
                 & ---             & $K^-$         &$0.050\pm0.004$  & $0.230\pm0.009$  & $434\pm30$    & $8\pm0.4$    &$10\pm0.3$                          & 8/11\\
                 & ---             & $p$           &$0.042\pm0.005$  & $0.214\pm0.010$  & $390\pm31$    & $9.8\pm0.5$  &$2\pm0.2$                           & 9/28\\
                 & ---             & $\bar p$      &$0.042\pm0.005$  & $0.214\pm0.010$  & $390\pm23$    & $9.8\pm0.5$  &$2\pm0.2$                           & 9/28\\
                 & ---             & $\phi$        &$0.050\pm0.006$  & $0.200\pm0.011$  & $312\pm21$    & $8\pm0.2$    &$0.007\pm0.0002$                    & 14/8\\
                 & ---             & $\Xi^-$       &$0.050\pm0.005$  & $0.187\pm0.010$  & $287\pm21$    & $8\pm0.3$    &$0.004\pm0.0002$                    &15/6\\
\hline
\end{tabular}%
\end{center}}
\end{table*}

\begin{figure*}[htbp]
\begin{center}
\includegraphics[width=15.0cm]{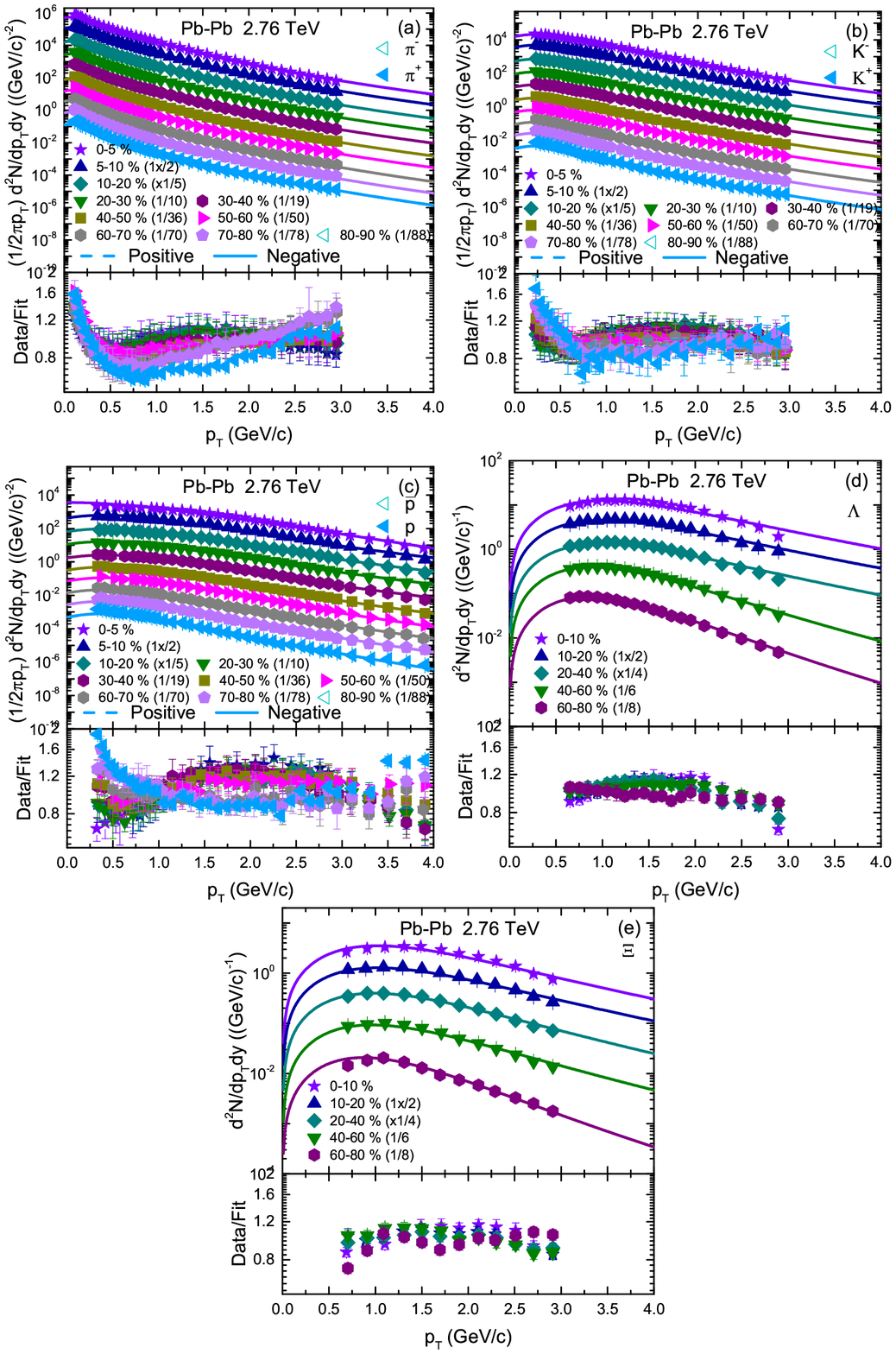}
\end{center}
\justifying\noindent {Fig. 4. The transverse momentum spectra of $\pi^+$, $\pi^-$, $K^+$, $K^-$,
$p$, $\bar p$, $\Lambda$ and $\Xi^-$ in Lead-Lead (Pb-Pb) collisions at 2.76 TeV in various centrality classes. The experimental data of ALICE Collaboration at LHC for $\pi^+$, $\pi^-$, $K^+$, $K^-$, $p$ and $\bar p$ are from ref. \cite{6c} at $|y|<$0.5 and for $\Lambda$ and $\Xi^-$ is taken from ref. \cite{7c} at $|y|<$0.5 mid-rapidity.}
\end{figure*}

\begin{figure*}[htbp]
\begin{center}
\includegraphics[width=15.0cm]{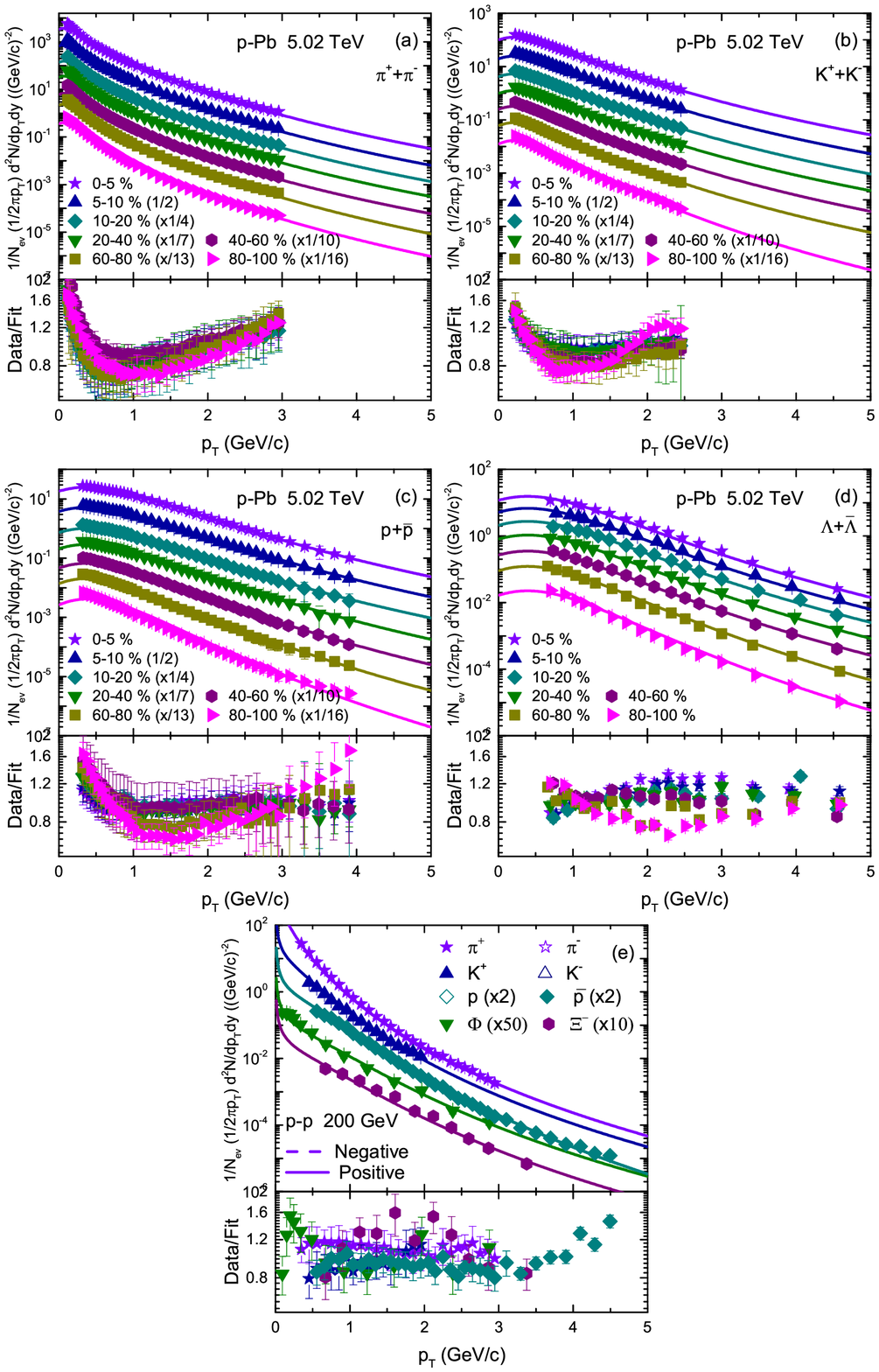}
\end{center}
\justifying\noindent {Fig. 5. Panel (a)-(d) show the transverse momentum spectra of $\pi^++\pi^-$, $K^++K^-$,
$p+\bar p$ and $\Lambda+\bar \Lambda$ in Lead-Lead (Pb-Pb) collisions at 5.02 TeV in various centrality classes in p-Pb collisions,and panel (e)
shows the transverse momentum spectra of $\pi^+$, $\pi^-$, $K^+$, $K^-$, $p$, $\bar p$, $\phi$ and $\Xi^-$ in $pp$ collisions at 200 GeV. The experimental data of BRAHMS Collaboration at LHC for
$\pi^++\pi^-$, $K^++K^-$, $p+\bar p$ and $\Lambda+\bar \Lambda$ in panels (a-e) are taken from ref. \cite{8c} at mid-rapidity at
$|y|<$0.5, while the data for $\pi^+$, $\pi^-$, $K^+$, $K^-$, $p$, $\bar p$, and the data for $\phi$ and $\Xi^-$ measured at STAR Collaboration are taken from ref. \cite{[9c,10c,11c} respectively at $|\eta|<$0.35 rapidity intervals for $\pi^+$, $\pi^-$, $K^+$, $K^-$, $p$, $\bar p$, and at $|y|<$0.5 for
$\phi$ and $\Xi^-$.}
\end{figure*}

\section{Results and discussion}

\subsection{Comparison with data}

Figure 1 shows the $p_{T}$ spectra ($1/N_{ev}$[(1/2$\pi$$m_T$) $d^2$$N$/$dyd$$m_T$]
or (1/2$\pi$$p_T$) $d^2$$N$/$dyd$$p_T$) of $\pi^+$, $\pi^-$, $K^+$, $K^-$, $p$,
$\bar p$, $\Lambda$ and $\Xi$ in Copper-Copper (Cu-Cu) collisions at 200 GeV in
different centrality bins. The centrality intervals for $\pi^+$, $\pi^-$, $K^+$,
$K^+$, $p$ and $\bar p$ are $0-10\%$, $10-30\%$, $30-50\%$ and $50-70\%$. The
centrality intervals for $\pi^+$ and $\pi^-$ $10-30\%$, $30-50\%$ and $50-70\%$
are scaled by 1/2, 1/4 and 1/6 respectively, while for $K^+$ and $K^-$ $30-50\%$
and $50-70\%$ are scaled by 0.3. For $p$ and $\bar p$ the centrality bins $10-30\%$,
$30-50\%$ and $50-70\%$ are scaled by 0.4, 0.3 and 0.2 respectively. The
centrality intervals for $\Lambda$ and $\Xi$ are $0-10\%$, $10-20\%$, $20-30\%$,
$30-40\%$ and $40-60\%$. $\pi^+$, $\pi^-$, $K^+$, $K^-$, $p$ and $\bar p$ are
measured at $y=0$, while for $\Lambda$ and $\Xi$ $|y|<$0.5. The symbols are used
to represent the experimental data of RHIC measured by the BRAHMS Collaboration, while
the curves are our fit results by using the Modified Hageodorn function with embeded
transverse flow i:e Eq. (3). The data/fit ratio is
given in the figure followed by each panel. It can be seen that Eq. (3) fits the data
approximately well. In some cases the fit in the $p_T$ range up to 0.5 GeV/c is not
good due to the resonance effect. The data for $\pi^+$, $\pi^-$, $K^+$, $K^-$, $p$,
$\bar p$ are taken from ref. \cite{1c}, while for $\Lambda$ and $\Xi$ is from ref. \cite{2c}.

Fig. 2 is similar to fig.1, but it shows the $p_T$ spectra of $\pi^+$, $K^+$,
$p$, $\Lambda$ and $\Xi^-$ in Au-Au collisions at 62.4 GeV. The centrality
is distributed in $0-5\%$, $0-10\%$, $10-20\%$, $20-30\%$, $30-40\%$, $40-50\%$, $50-60\%$,
$60-70\%$ and $70-80\%$ intervals for $\pi^+$, $K^+$ and $p$ and the centrality bins
$0-10\%$, $10-20\%$, $20-30\%$, $30-40\%$, $40-50\%$, $50-60\%$, $60-70\%$ and $70-80\%$ are
scaled by 1/4, 1/15, 1/50, 1/80, 1/500, 1/1300, 1/2700 and 1/5400 respectively. $\Lambda$
is distributed into $0-5\%$, $0-10\%$, $10-20\%$, $20-30\%$, $30-40\%$, $40-60\%$ and $60-80\%$
centrality intervals, while the centrality distribution for $\Xi^-$ is $0-5\%$,
$0-10\%$, $10-20\%$, $20-30\%$, $30-40\%$ and $40-60\%$. One can see that the measurement of the STAR
Collaboration is well fitted by the modified Hagedorn function. The data/fit ratio is
followed in each panel in the figure. The data for $\pi^+$, $K^+$ and
$p$ are taken from ref. \cite{3c}, while for $\Lambda$ and $\Xi$ is taken from ref. \cite{4c}.

The $p_T$ spectra of the given particles are demonstrated in fig.3 and 4 in Deuteron-Gold
(d-Au) collisions at 200 GeV and in Lead-Lead (Pb-Pb) collisions at 2.76 TeV respectively.
In d-Au collisions $\pi^+$, $\pi^-$, $K^+$, $K^-$, $p$ and $\bar p$, and in Pb-Pb collisions $\pi^+$, $K^+$, $p$, $\Lambda$ and $\Xi$ are analyzed. The centrality in d-Au is
distributed in various centrality classes such that $0-20\%$, $20-40\%$, $40-60\%$ and $60-88\%$.
However the centrality for $\pi^+$, $\pi^-$, $K^+$, $K^-$, $p$ and $\bar p$ in Pb-Pb collisions
is distributed in $0-5\%$, $0-10\%$, $10-20\%$, $20-30\%$, $30-40\%$, $40-50\%$, $50-60\%$,
$60-70\%$, $70-80\%$ and $80-90\%$, and except $0-5\%$, the other centrality bins are scaled by
1/2, 1/5, 1/10, 1/19, 1/36, 1/50, 1/70, 1/78 and 1/88 respectively. The centrality bins for
$\Lambda$ and $\Xi$ are $0-10\%$, $10-20\%$, $20-40\%$, $40-60\%$ and $60-80\%$ and except $0-10\%$,
the remaining centrality class are scaled by 1/2, 1/4, 1/6 and 1/8 respectively.
The well fitting results of the model to the experimental data of ALICE collaboration can be seen, and the data/fit ratio
are followed in each panel. The data in fig. 3 are taken from ref. \cite{5c}, while in fig.4 for $\pi^+$,
$\pi^-$, $K^+$, $K^-$, $p$ and $\bar p$ is taken from ref. \cite{6c}, and $\Lambda$ and $\Xi^-$ are taken
from ref. \cite{7c}.

Fig.5 shows the $p_T$ of the given particles in proton-Lead (p-Pb) and proton-proton ($pp$) collisions.
Panel (a)-(d) shows the $p_T$ spectra of the particles in p-Pb collisions in different centrality intervals
at 5.02 TeV, and panel (e) represent the $p_T$ spectra of the particles in $pp$ collisions at 200 GeV.
Various centrality bins for $\pi^++\pi^-$, $K^++K^-$, $p+\bar p$ and $\Lambda+\bar \Lambda$ are $0-5\%$,
$0-10\%$, $10-20\%$, $20-40\%$, $40-60\%$, $60-80\%$, and $80-100\%$, and $0-10\%$, $10-20\%$, $20-40\%$,
$40-60\%$, $60-80\%$, and $80-100\%$ in panel (a)-(c) are scaled by 1/2, 1/4, 1/7, 1/10, 1/13 and 1/16
respectively. Panel (e) represents the $p_T$ spectra of $\pi^+$, $\pi^-$, $K^+$, $K^-$, $p$, $\bar p$,
$\phi$ and $\Xi^-$, and the spectra of $p$ and $\bar p$, $\phi$ and $\Xi^-$ are scaled by 2, 50 and 10
respectively. One can see that the modified Hagedorn function fits the experimental data well. The results
of the data/fit ratio are given in figure followed by each panel. The data for  $\pi^++\pi^-$, $K^++K^-$,
$p+\bar p$ and $\Lambda+\bar \Lambda$ in panels (a-e) is taken from ref. \cite{8c}, while the data for $\pi^+$,
$\pi^-$, $K^+$, $K^-$, $p$, $\bar p$, and $\phi$ and $\Xi^-$ are taken from ref. \cite{9c,10c,11c} respectively.

\subsection{Tendencies of parameters}

In order to study the tendency of parameters, the dependences of $T_0$ on centrality in
(a)-(e) and $T_0$ on $m_0$ in (f) with different symbols are
displayed in Figure 6. The values of the two parameters are cited from Table 1. Each panel
represent the results from different collisions. The symbols in the figure are used in order
to represent different particles. The parameters trend from left to right shows the behavior
of $T_0$ from central to peripheral collisions from panel (a)-(e) while panel (f) shows its
behavior with increasing mass of the particle. In panels (a)-(e) one can see that $T_0$ for all the particles is larger in central
collisions and it decrease as the centrality decrease due the reason that more energy is deposited
to the system in central collisions due to the involvement of large number of participants which
decrease towards periphery. In addition, from panel (a)-(f), $T_0$ is larger for the strange particles
and the separate decoupling of the strange and non-strange particles is observed. The larger $T_0$ for
the strange particles is due to their smaller cross-section interaction, and according to kinematics,
the reactions with lower cross-section is expected to be switched-off at higher temperatures/densities
or early in time than the reactions with higher cross-sections. We also believe that the
charm particles may decouple earlier than the strange particles and it is possible that a series of freeze-outs
correspond to particular reaction channels \cite{12c}, but it is a regret that we don't have the data for charm particles.
Furthermore, one can see that $T_0$ is larger at LHC than that of RHIC, and even at RHIC $pp$ is the smallest collisions
system and the observed $T_0$ in it is very smaller than the rest which is a clear evidence of $T_0$ on the size of the
interacting system. From the above observation, one thing
is clear that there is a strong dependence of $T_0$ on the collision energy because Pb-Pb and p-Pb collisions
are the largest systems and they have larger $T_0$, and $pp$ is the smallest collision system among them and
has smaller vale for $T_0$. However Cu-Cu, Au-Au and d-Au at RHIC, and p-Pb and Pb-Pb at LHC are different systems and
have almost the same values for $T_0$ respectively, and this is due to the effect of collision energy. So we can conclude that $T_0$ is
dependent on the size of the collision system as well as the collision energy.

We would like to point out that we have observed two kinetic freeze-out scenarios which is agreement with \cite{13c}, but in contrast with our recent work \cite{13ca}. In fact, there are different freeze-out scenarios in literature which include one, two, three or multiple kinetic freeze-out scenarios. No doubt, the process of high energy collisions is complex and it is very interesting to see whether one, two, or multiple kinetic freeze-out scenarios exist in it. Due to the fact that different scenarios are reported in different literature, it is an open question up
to now.

\begin{figure*}[htbp]
\begin{center}
\includegraphics[width=11.0cm]{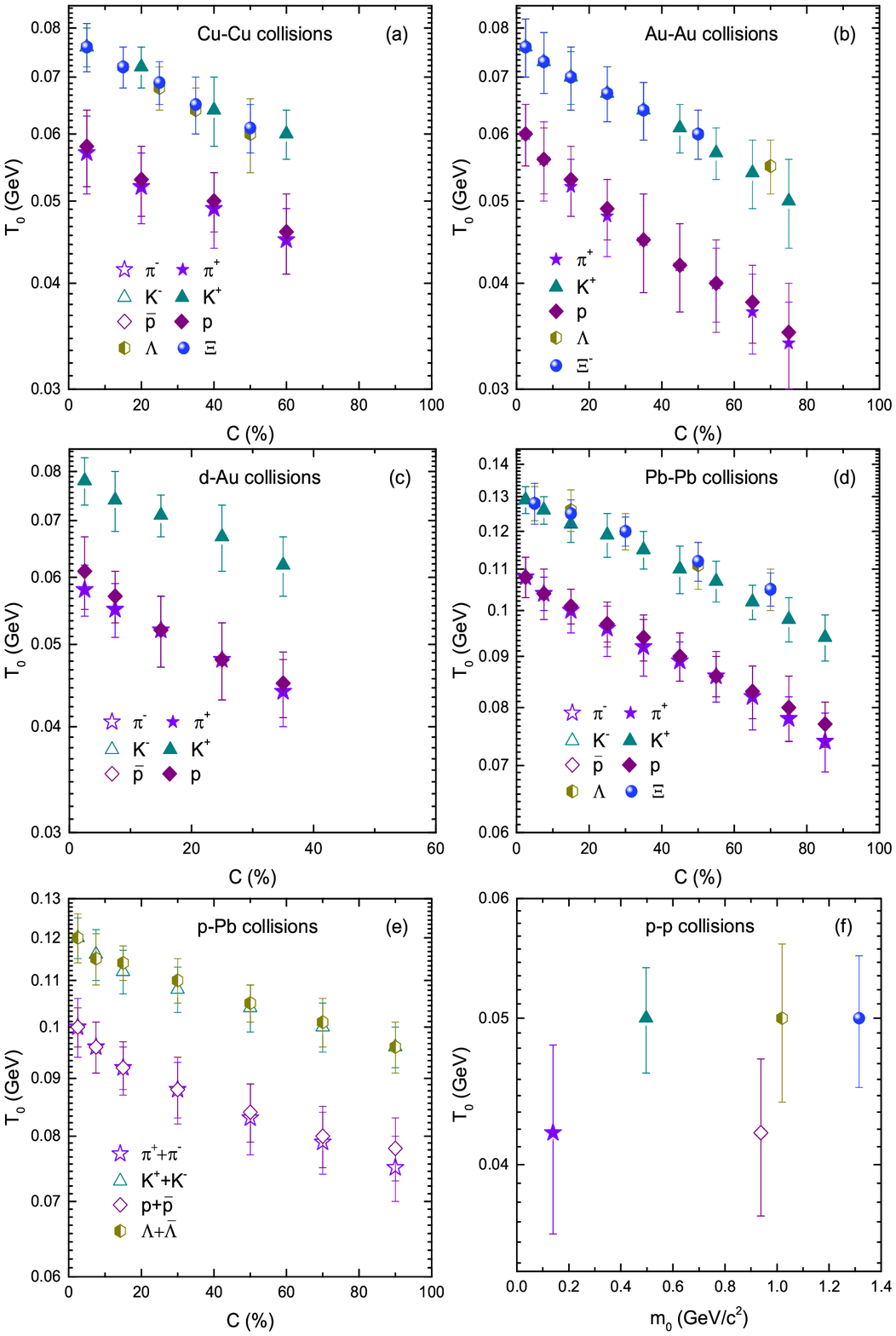}
\end{center}
\justifying\noindent {Fig. 6. Dependence of $T_0$ on centrality in panel (a)-(e), and dependence of $T_0$ on $m_0$ in panel (f).}
\end{figure*}

Fig. 7 shows similar to the fig. 6, but it shows the dependence of $\beta_T$ on centrality from panel (a)-(e), while panel (f)
shows the dependence of $\beta_T$ on $m_0$. One can see that $\beta_T$ remains invariant as we go from central to peripheral collision
due to the reason that collective behavior from central to peripheral collisions does not change. In addition, $\beta_T$ from panel
(a)-(f)is larger for the lighter particles and it decrease for the heavier particles and it is a natural hydrodynamical behavior
because heavier particles are left behind in the system. We also noticed that like $T_0$, $\beta_T$ in
Cu-Cu, Au-Au and d-Au at RHIC, and Pb-Pb and p-Pb at LHC are close to each other due to the effect of their dependence on both
the collision cross-section and collision energy.
\begin{figure*}[htbp]
\begin{center}
\includegraphics[width=11.0cm]{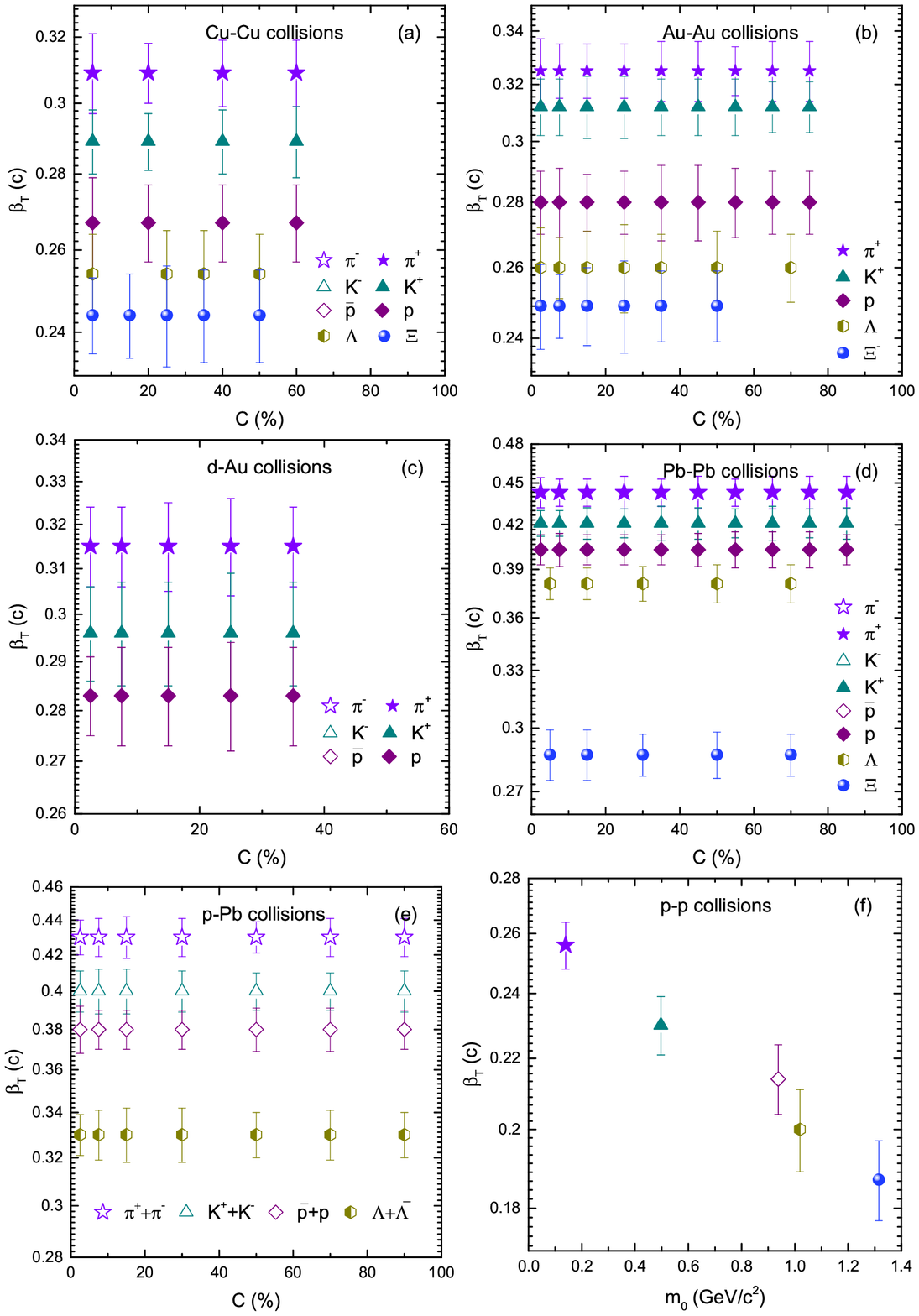}
\end{center}
\justifying\noindent {Fig. 7. Dependence of $\beta_T$ on centrality in panel (a)-(e), and dependence of $\beta_T$ on $m_0$ in panel (f).}
\end{figure*}

To check the dependence of $V$ on centrality (mass), fig. 8 is represented. It is similar to that of fig. 6 and fig. 7. One can see that
$V$ decreases from central to peripheral collisions in panels (a)-(e) due to fact that the participant nucleons decreases as we go from central
to peripheral collisions depending on the interaction volume. The system which contains more participants reaches to equilibrium quickly because
there are large number of secondary collisions by the re-scattering of partons in central collisions, and the system goes away from equilibrium
states as it goes from central to peripheral collision due to reason that the number of participants in the system decreases. Furthermore, we observed
that $V$ from panel (a)-(f) is dependent of $m_0$. Heavier the particle is, smaller is the value of $V$, which reveals a volume differential freeze-out
scenario, and may indicate that there are different freeze-out surfaces for different particles. Like $T_0$ and $\beta_T$, $V$ at LHC is larger than
at RHIC and in $pp$ collisions $V$ is the smallest among all these systems, nucleus-nucleus collision systems at
RHIC and LHC have nearly same $V$ respectively and this is due to their energy difference.
\begin{figure*}[htbp]
\begin{center}
\includegraphics[width=11.0cm]{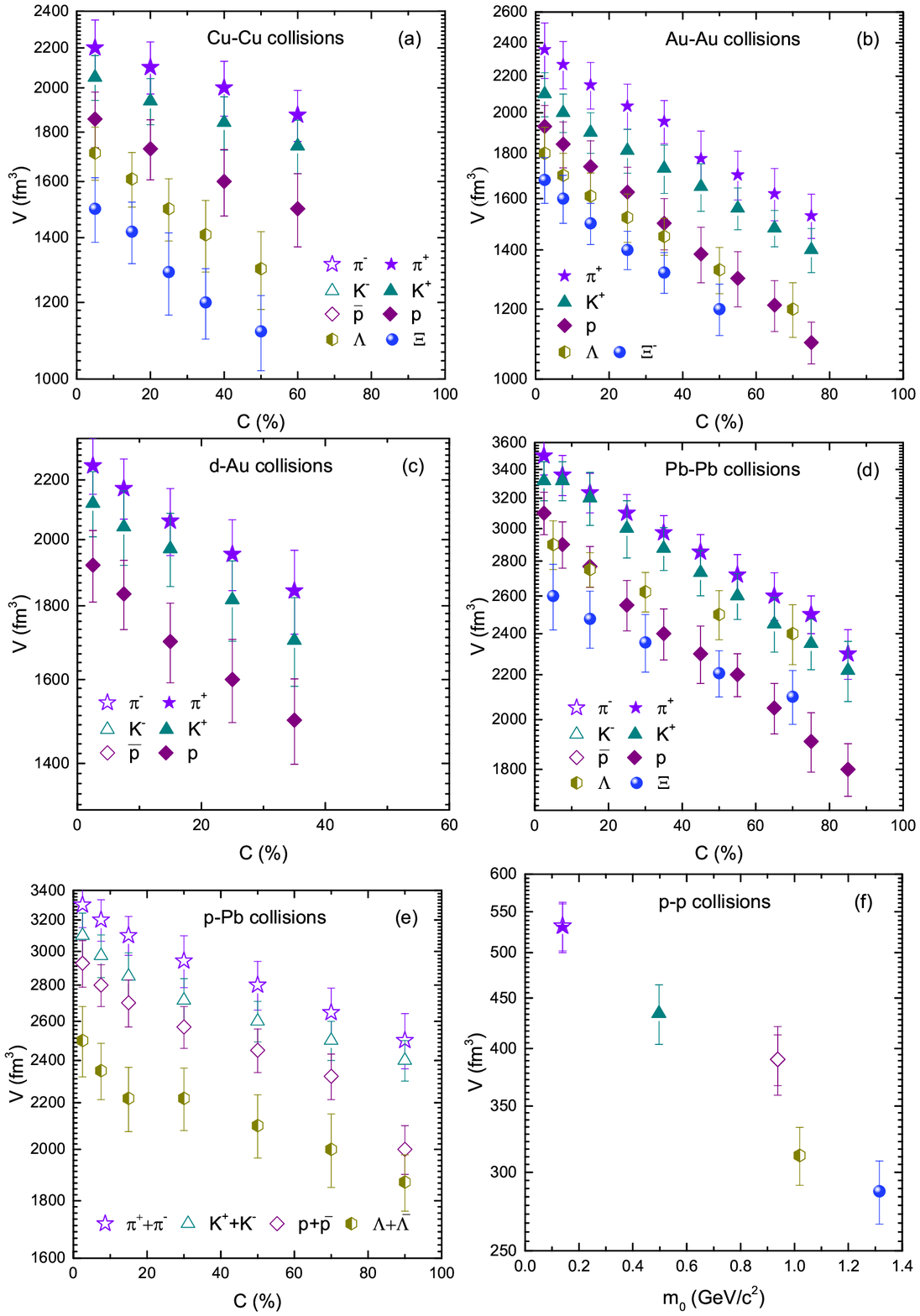}
\end{center}
\justifying\noindent {Fig. 8. Dependence of $V$ on centrality in panel (a)-(e), and dependence of $V$ on $m_0$ in panel (f).}
\end{figure*}

Fig. 9 represents the behavior of the parameter $n$. $n$=1/(q-1), and q is the entropy index, which is referred to
equilibrium degree. In general, $q$=1 corresponds to equilibrium state, where $q$$>$1 corresponds to non-equilibrium state.
Smaller $q$ corresponds to large $n$. One can see that the parameter $n$ is larger in almost all the cases which means that
the system is in equilibrium state, and also it remains constant in most cases from central to peripheral collision except
for proton that increases from central to peripheral collisions which is not understood by us.
\begin{figure*}[htbp]
\begin{center}
\includegraphics[width=11.0cm]{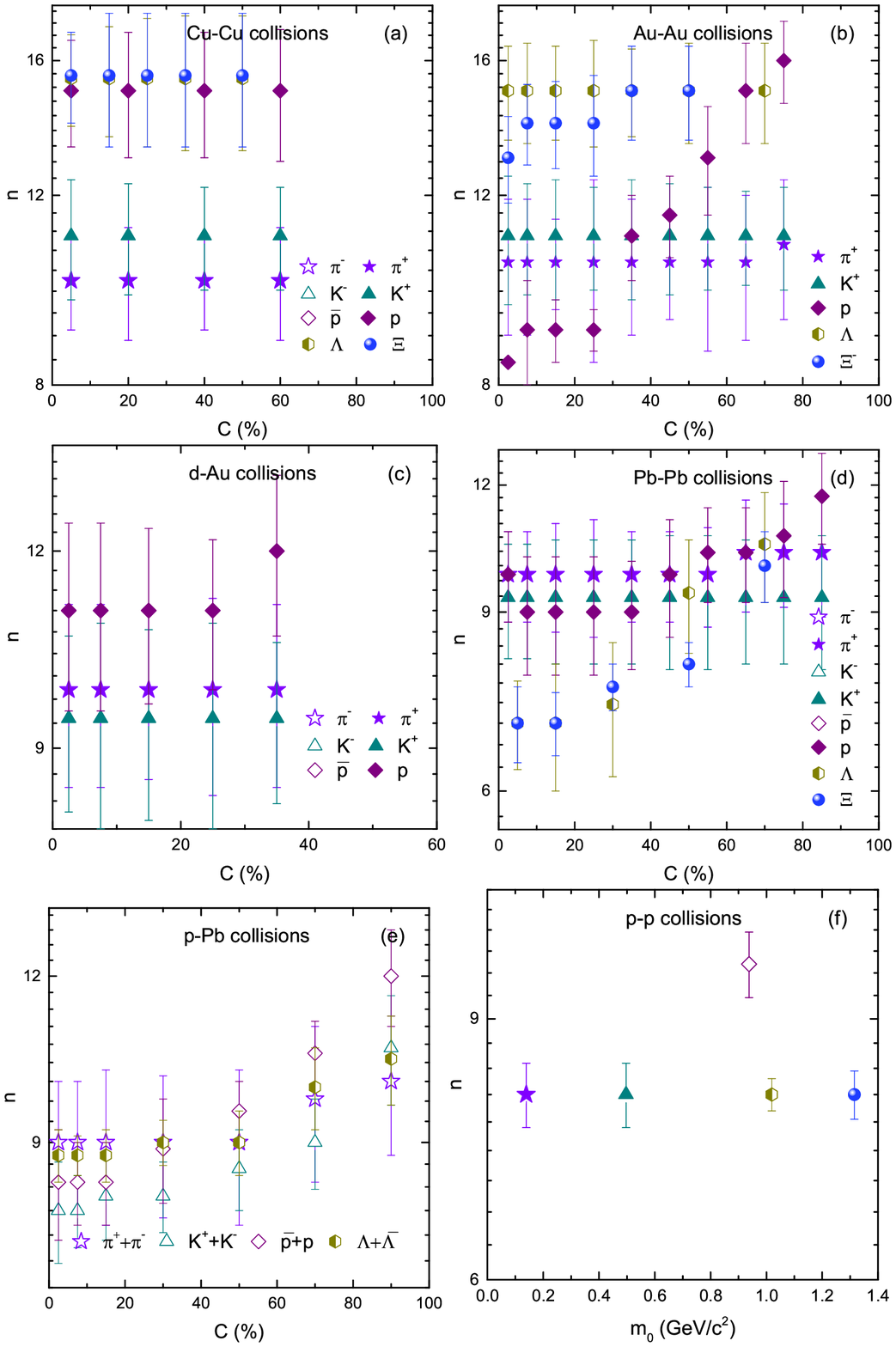}
\end{center}
\justifying\noindent {Fig. 9. Dependence of $n$ on centrality in panel (a)-(e), and dependence of $n$ on $m_0$ in panel (f).}
\end{figure*}

\begin{figure*}[htbp]
\begin{center}
\includegraphics[width=11.0cm]{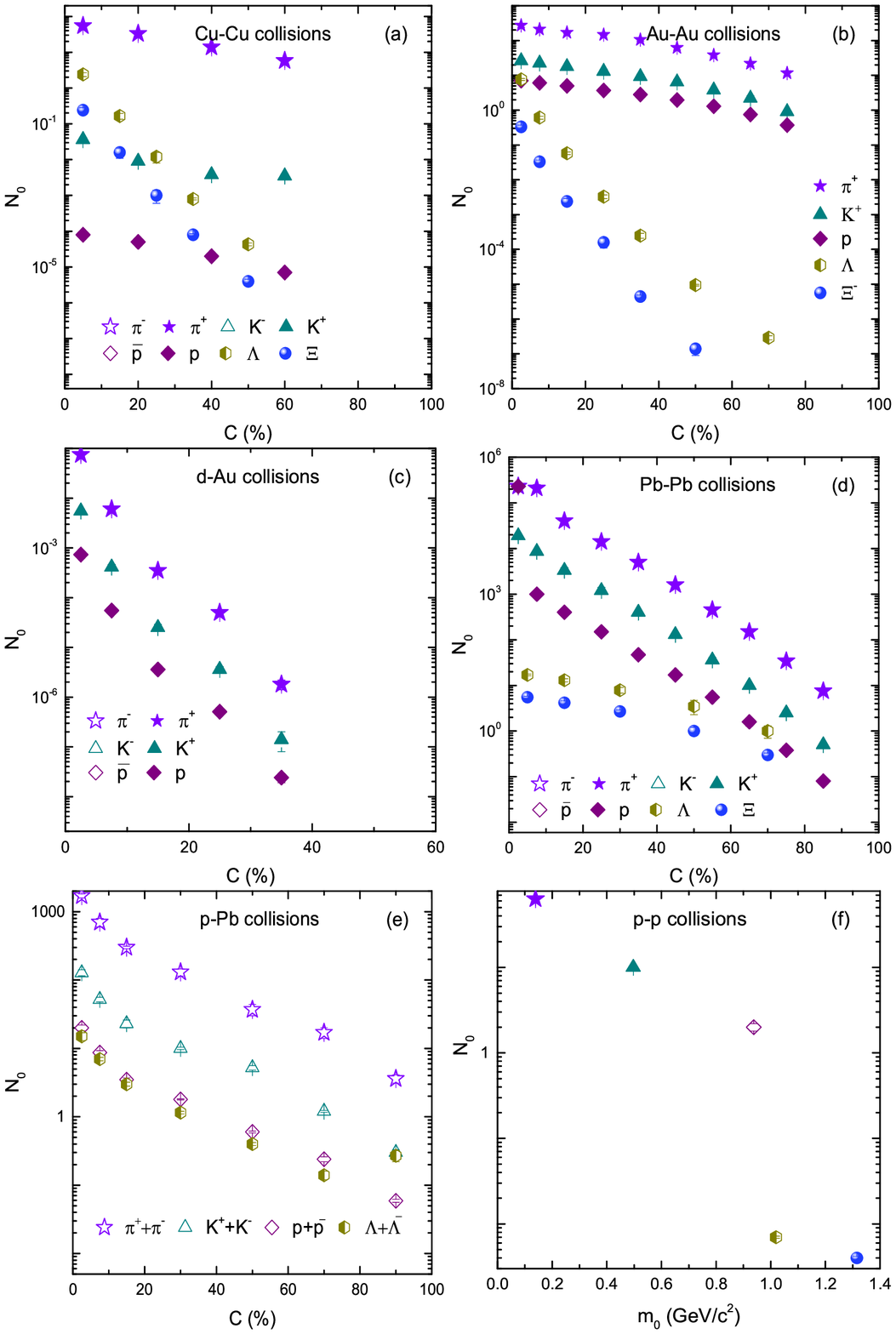}
\end{center}
\justifying\noindent {Fig. 10. Dependence of $N_0$ on centrality in panel (a)-(e), and dependence of $N_0$ on $m_0$ in panel (f).}
\end{figure*}

The dependence of the parameter $N_0$ on centrality ($m_0$). $N_0$ is a normalization constant but it has its significance too. It
reflects the multiplicity. One can see that the parameter $N_0$ decreases from central to peripheral collisions which indicates
that the multiplicity decreases as the system goes from central to peripheral collisions. It can also be seen that $N_0$ is
dependent on the size of the interacting system. As we can see that in Au-Au collisions it is larger than Cu-Cu and d-Au collisions,
and in Pb-Pb collisions it has the highest value followed by p-Pb collisions. In $pp$ collisions $N_0$ is the smallest due to being it is
the smallest system among them. This shows that the multiplicity depends on the size of the interacting system and also have the collision
energy dependence as the collision energy of the AA collisions at RHIC and LHC are different.

\begin{figure*}[htbp]
\begin{center}
\includegraphics[width=11.0cm]{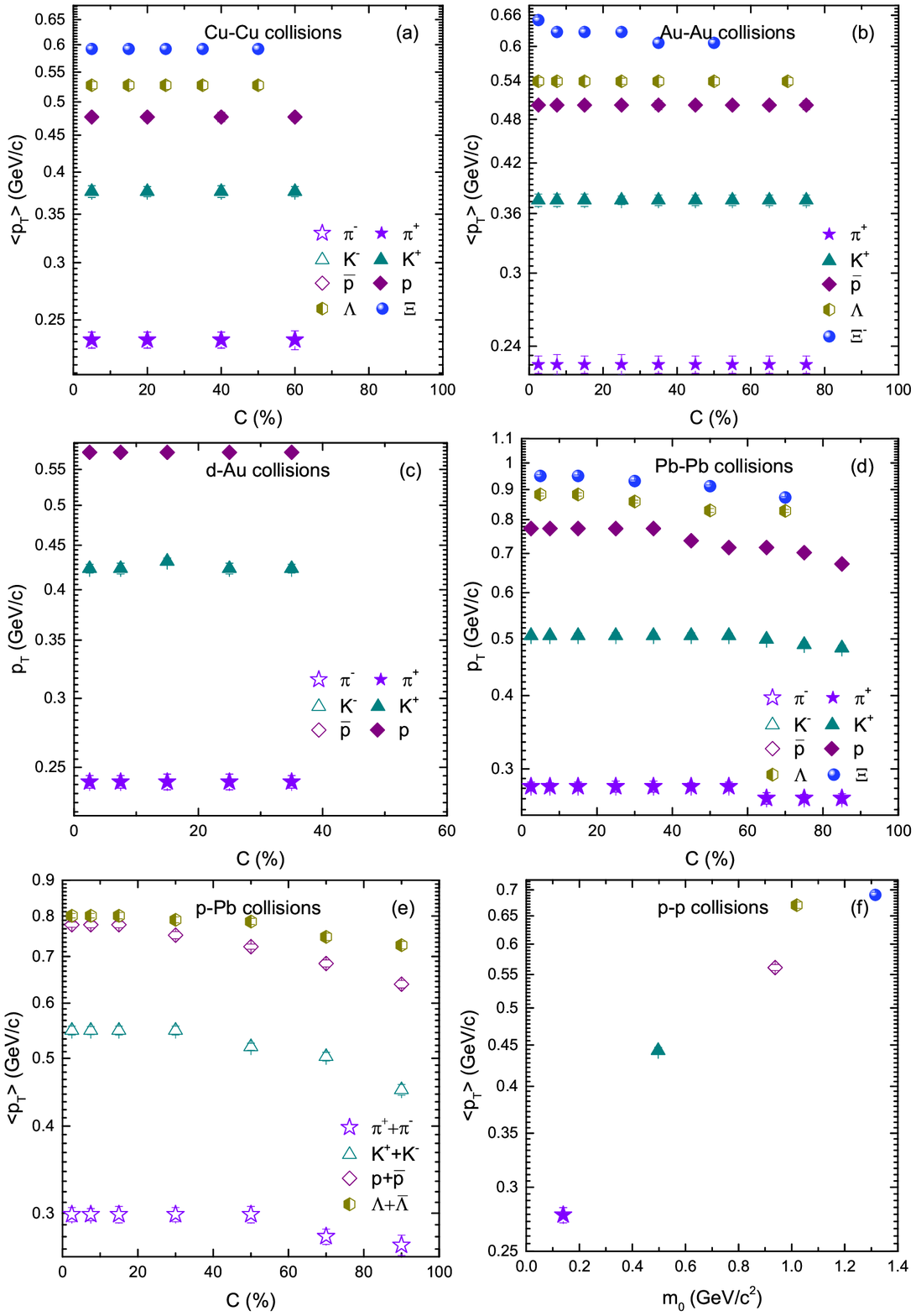}
\end{center}
\justifying\noindent {Fig. 11. Dependence of $<p_T>$ on centrality in panel (a)-(e), and dependence of $<p_T>$ on $m_0$ in panel (f).}
\end{figure*}

\begin{figure*}[htbp]
\begin{center}
\includegraphics[width=11.0cm]{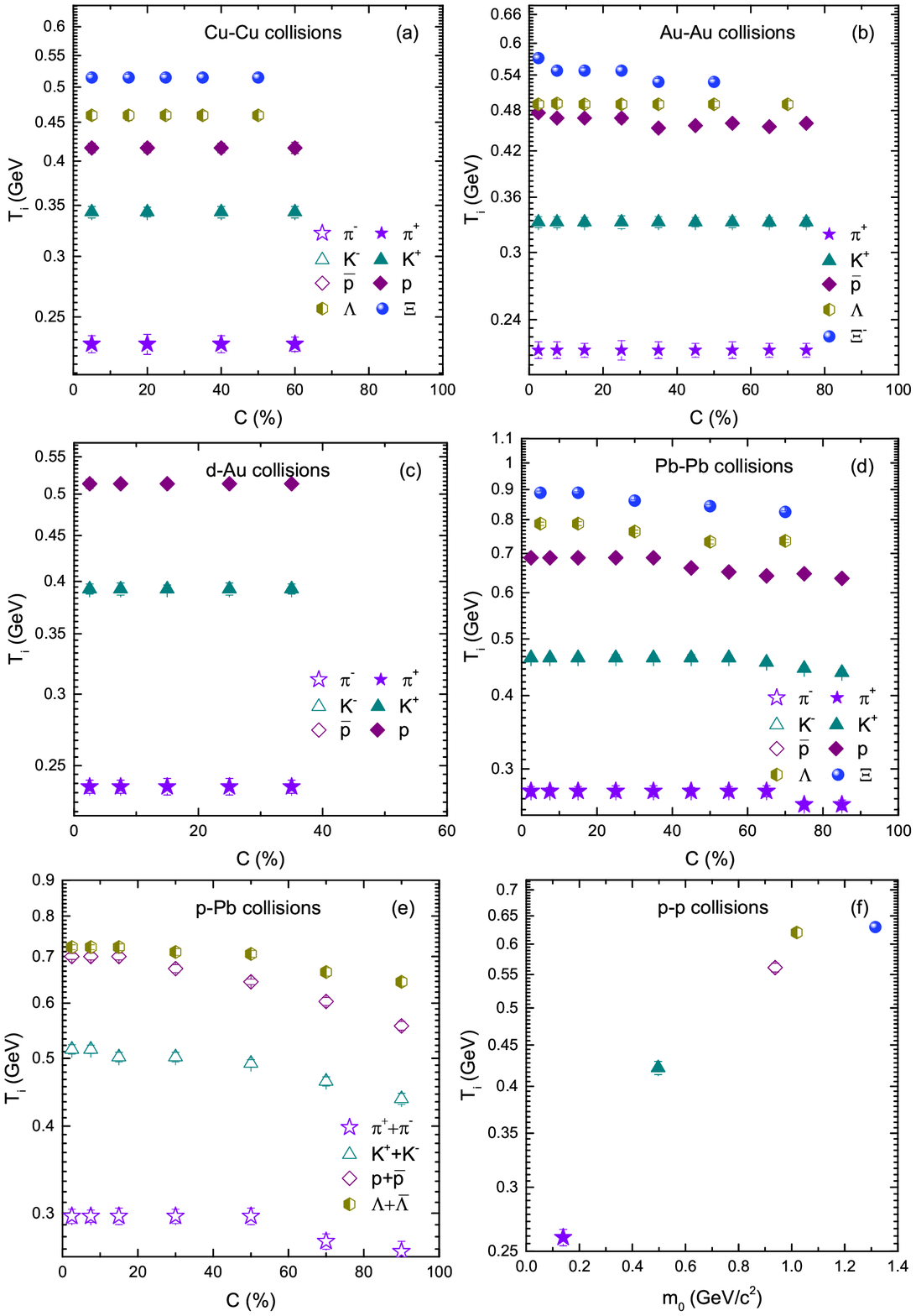}
\end{center}
\justifying\noindent {Fig. 12. Dependence of $T_i$ on centrality in panel (a)-(e), and dependence of $T_i$ on $m_0$ in panel (f).}
\end{figure*}
The mean transverse momentum ($<p_T>$) dependence on centrality ($m_0$) is demonstrated in fig. 11. One can see that $<p_T>$
decrease as the system goes from central to peripheral collisions and this is due to the fact that the system gains large
momentum (energy) in central collisions where further multiple scattering happens, which decrease when the system goes towards
periphery. Like $T_0$ and $\beta_T$, $<p_T>$ also depends on the size of the interacting system and also there is an effect
on the behavior of collision energy. We also analyze root-mean-square $p_T$ ($\sqrt{<p^2_T>}$) over $\sqrt2$ ($\sqrt{<p^2_T>}$/$\sqrt2$)
in fig. 12. and showed its behavior with changing the centrality ($m_0$). ($\sqrt{<p^2_T>}$/$\sqrt2$) represents the initial
temperature ($T_i$) of the interacting system according to the string percolation model \cite{14c,15c,16c}. It can be obviously seen that the
initial temperature is larger in central collisions, however it decrease from central to peripheral collision systems, and similar to
$T_0$ it depends on the size of the interacting system and is also effected by the collision energy. Moreover, we have observed that
the initial temperature depends on mass of the particle. The more massive the particle is, the larger is the initial temperature.

We would like to point out that the initial temperature depends on mass of the particle, and the more heavier the particle is, the
larger the initial temperature, however the kinetic freeze-out temperature depends on the cross-section interaction of the particles.
The larger the cross-section of the particle is, the smaller the value of kinetic freeze-out temperature. The difference in the two
temperatures is due to the reason that they occur at two different process in the system evolution. Furthermore, it is also possible
that the centrality dependence of the two types of temperatures is also different if one gets the increasing trend from central
to peripheral for $T_0$ which is observed in our recent work \cite{17c, 18c}. In fact the trend dependence of $T_0$ is also an
open question in high energy collisions because different literature give different trend.

Before going to the summary and conclusions, we would like to point out that the initial temperature is observed to be larger than
the kinetic freeze-out temperature. The former is followed by the chemical freeze-out temperature which can be expresses as
\begin{align}
T_{ch} = \frac{T_{\lim}}{1+\exp[2.60-\ln(\sqrt{s_{NN}})/0.45]}
\end{align}
where $T_{lim}$=0.1584 GeV. The chemical freeze-out temperature is followed by the effective temperature and then the kinetic
freeze-out temperature and this order is in agreement with the order of time evolution of the interacting system.

\section{Summary and Conclusions}

We summarize here our main observations and conclusions.

(a) The transverse momentum spectra of strange and non-strange hadrons produced in
Cu-Cu, Au-Au, d-Au, Pb-Pb, p-Pb and $pp$ collisions have been studied by the modified Hagedorn model.
The results are well in agreement with the experimental data BRAHM, STAR, PHENIX and ALICE Collaborations
at RHIC and LHC.

(b) $T_0$ and $T_i$ are larger in central collisions and they decrease from central to peripheral collisions due to
the decrease in the participant nucleons towards periphery which results in lower degree of excitation of the
system in the peripheral collisions. $<p_T>$ also decrease towards periphery due to the reason that
the energy (momentum) transfer becomes lower in the system from central to peripheral collisions. In addition $V$
is larger in central collisions and it decrease towards periphery due to the reason of decreasing of partons re-scattering towards periphery.

(c) $\beta_T$ depends on the mass of the particle. Heavier the particle is, smaller is the value of $\beta_T$ is. $\beta_T$
remains unchanged from central to peripheral collisions because the collective flow from central to peripheral collisions
does not change.

(d) The parameter $N_0$ represents the multiplicity and it decrease from central to peripheral collisions.

(e) $T_0$ depends on the cross-section interaction of the particle and therefore the strange and non-strange
particles have separate freeze-out and it reveals the scenario of two kinetic freeze-out temperature, while the initial
temperature depends on the mass of the particle and this is due to the reason that both of them occurs at different stages in the
evolution system.

(f) $V$ depends on the mass of the particle. The heavier particle has smaller $V$ which reveals the volume differential
freeze-out scenario and indicates a separate freeze-out surface for each particle.

(g) $T_0$, $T_i$, $\beta_T$, $V$, $<p_T>$ and $N_0$ are dependent on the size of the interacting system because all
these parameters have larger values at LHC collision systems than at RHIC collision systems and in $pp$ collisions it has the lowest
values. The mentioned parameters in Cu-Cu, d-Au, and Au-Au are different collision systems with different energies (200 GeV, 62.4 GeV and 200 GeV for Cu-Cu, Au-Au and d-Au respectively), but the values of the parameters obsered in these systems are nearly equal, due to their dependence on both the collision energy and collision cross-section. Similar behavior is observed in p-Pb and Pb-Pb collisions.
\\
\\
\\
{\bf Acknowledgments}

This research was funded by the National Natural Science Foundation of China grant
number 11875052, 11575190, and 11135011.
\\
\\
{\bf Author Contributions} All authors listed have made a
substantial, direct, and intellectual contribution to the work and
approved it for publication.
\\
\\
{\bf Data Availability Statement} This manuscript has no
associated data or the data will not be deposited. [Authors'
comment: The data used to support the findings of this study are
included within the article and are cited at relevant places
within the text as references.]
\\
\\
{\bf Compliance with Ethical Standards}
\\
\\
{\bf Ethical Approval} The authors declare that they are in
compliance with ethical standards regarding the content of this
paper.
\\
\\
{\bf Disclosure} The funding agencies have no role in the design
of the study; in the collection, analysis, or interpretation of
the data; in the writing of the manuscript, or in the decision to
publish the results.
\\
\\
{\bf Conflict of Interest} The authors declare that there are no
conflicts of interest regarding the publication of this paper.
\\
\\

\end{document}